\documentclass[sigconf]{acmart}
\AtBeginDocument{%
  }

\setcopyright{acmlicensed}
\copyrightyear{2018}
\acmYear{2018}
\acmDOI{XXXXXXX.XXXXXXX}

\acmConference[Conference acronym 'XX]{Make sure to enter the correct
  conference title from your rights confirmation emai}{June 03--05,
  2018}{Woodstock, NY}
\acmISBN{978-1-4503-XXXX-X/18/06}
\usepackage{xcolor}
\usepackage{fancybox}
\usepackage{CJK}

\usepackage{listings}

\copyrightyear{2025}
\acmYear{2025}
\setcopyright{acmlicensed}\acmConference[CHI '25]{CHI Conference on Human Factors in Computing Systems}{April 26-May 1, 2025}{Yokohama, Japan}
\acmBooktitle{CHI Conference on Human Factors in Computing Systems (CHI '25), April 26-May 1, 2025, Yokohama, Japan}
\acmDOI{10.1145/3706598.3714115}
\acmISBN{979-8-4007-1394-1/25/04}

\newcommand{\revision}[1]{{\color{black} #1}}
\begin{document}
%\tcbset{colback=white, colframe=black, boxrule=0.5mm, arc=4mm, outer arc=4mm}

\title{DynEx: Dynamic Code Synthesis with Structured Design Exploration for Accelerated Exploratory Programming}


% \author{Jenny Ma, Karthik Sreedhar, Vivian Liu, Pedro A. Perez, Sitong Wang, Riya Sahni, Lydia B. Chilton}
% \affiliation{
%   \institution{Columbia University}
%   % \city{New York}
%   % \state{New York}
%   % \country{USA}
% }
% \email{jenny.ma@columbia.edu} \email{chilton@cs.columbia.edu}

\author{Jenny Ma, Karthik Sreedhar, Vivian Liu, Pedro A. Perez, Sitong Wang, Riya Sahni, Lydia B. Chilton}
\affiliation{
  \institution{Columbia University}
    \city{New York}
  \state{New York}
  \country{USA}\\
  \href{mailto:jenny.ma@columbia.edu, chilton@cs.columbia.edu}{jenny.ma@columbia.edu, chilton@cs.columbia.edu}
}
\renewcommand{\shortauthors}{Ma et al.}

\begin{abstract}
Recent advancements in large language models have significantly expedited the process of generating front-end code.
This allows users to rapidly prototype user interfaces and ideate through code, a process known as exploratory programming.
However, existing LLM code generation tools focus more on technical implementation details rather than finding the right design given a particular problem.
We present DynEx, an LLM-based method for design exploration in accelerated exploratory programming. 
DynEx introduces a technique to explore the design space through a structured Design Matrix before creating the prototype with a modular, stepwise approach to LLM code generation. Code is generated sequentially, and users can test and approve each step before moving onto the next.
A user study of 10 experts found that DynEx increased design exploration and enabled the creation of more complex and varied prototypes compared to a Claude Artifact baseline. 
We conclude with a discussion of the implications of design exploration for exploratory programming. 

\end{abstract}

\begin{CCSXML}
<ccs2012>
   <concept>
       <concept_id>10003120.10003121.10003129.10011756</concept_id>
       <concept_desc>Human-centered computing~User interface programming</concept_desc>
       <concept_significance>500</concept_significance>
       </concept>
 </ccs2012>
\end{CCSXML}

\ccsdesc[500]{Human-centered computing~User interface programming}

\keywords{code synthesis, exploratory programming, design exploration, design matrix, user interface, prototyping}

\begin{teaserfigure}
  \includegraphics[width=\textwidth]{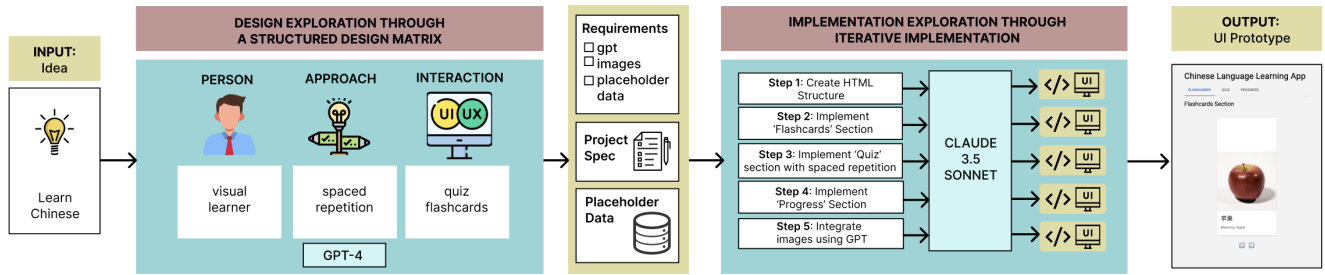}
  \caption{\textbf{DynEx} is an LLM-based system for exploratory programming. It guides users through a design space using a structured Design Matrix. Based on the Design Matrix content, DynEx brainstorms the necessary requirements needed to build the UI, generates a detailed project specification to translate the abstract concepts into an implementation plan, and generates data to better prototype the application. DynEx then breaks down the project into tasks, which users can iteratively implement until they create a working prototype. Users can incrementally add features until they are satisfied, and also create a variety of new prototypes using the system. }
  \Description{The figure depicts the workflow of DynEx. Users first input an application idea into the system - to learn Chinese. DynEx guides users through a design space using a structured Design Matrix, with dimensions of Person, Approach, and Interaction. For this example, the Person is a visual learner, the Approach is spaced repetition, and the Interaction is quiz flashcards. Based on the Design Matrix content, DynEx brainstorms the necessary requirements needed to build the UI, generates a detailed project spec to translate the abstract concepts into implementation goals, and generates synthetic placeholder data to better prototype the application. DynEx then breaks down the project into tasks, which users can iteratively implement until they create a working prototype. Here, DynEx breaks it down into 5 tasks: 1) Create the HTML Structure, 2) Implement 'Flashcards' Section, 3) Implement 'Quiz' section with spaced repetition, 4) Implement 'Progress' section, and 5) Integrate images using GPT. Users can incrementally add features until they are satisfied, and also can go create a variety of new prototypes using the system. The final output is a UI prototype that users can interact with - a Chinese learning app with images}
  \label{fig:teaser}
\end{teaserfigure}

\received{20 February 2007}
\received[revised]{12 March 2009}
\received[accepted]{5 June 2009}

\maketitle

\section{Introduction}
Recently released large language models (LLMs) have shown remarkable capabilities in generating code, particularly front-end code ~\cite{si2024design2codefarautomatingfrontend}.
% Recent code-synthesis advancements in large language models (LLMs) have significantly accelerated the process of generating code, particularly front-end code ~\cite{si2024design2codefarautomatingfrontend}. 
This allows users to get ideas off the ground faster by building code-based user-interface (UI) prototypes and testing them directly in real world contexts, a process known as exploratory programming.
Exploratory programming is crucial for experimental projects where real design testing is more productive than upfront specifications. 
Many interactions are difficult to simulate in low-fidelity prototyping mediums, particularly when testing data-driven applications~\cite{8103446}. Prototyping is crucial in these cases to truly test ideas.
LLM code-generation abilities present a unique opportunity to accelerate exploratory programming and ideate through code.

% and thus require programming in the prototyping process 
%Even high-fidelity approaches like Figma have their limitations at modeling the interactive experience, because high-fidelity prototyping takes time and resources to emulate user experience across frames of UI. Time and resource costs also grow as prototyping gets more high-fidelity ~\cite{tiong2019economies}. 

When prototyping, however, translating abstract concepts into concrete implementations is a challenging process ~\cite{tranquillo2015coding}.
There is a gap in going from an idea to a working solution because initial ideas often lack clear boundaries, structured organization, and concrete details. 
Converting an initial idea into a feasible design requires a detailed consideration of multiple aspects: users must fully consider the problem being solved, the target user of the application, the approach or methodology, and the interaction paradigm or user experience
%because user interfaces are most effective when personalized towards the intended user 
~\cite{arazy2015personalityzation}.
Each of these elements must be defined before moving forward with implementation, making the transition from idea to design a complex and multifaceted process.
%Moreover, implementation itself is costly and time-consuming, as programming involves navigating between high-level concepts and detailed technical execution.

Existing tools that leverage LLM code-generation capabilities, such as GPTPilot ~\cite{gpt_pilot} and OpenHands (formerly OpenDevin) ~\cite{wang2024opendevin}, predominantly focus on code development rather than creating an application that considers the end-user experience.
GPTPilot prompts the user to consider implementation-level details, such as which packages to use, rather than broader aspects of the user’s ideas: who the application is for, what problem the application is trying to solve, and what the core methodology is for guiding the solution. Claude Artifact is more suitable for exploration as it enables users to chat with the LLM, but its fundamentally unstructured approach can also be limiting. Users easily become fixated on implementation over exploration, appending features onto prototypes linearly rather than using the chatbot to laterally exploring their problem space ~\cite{claude_3_5_sonnet}.
Ultimately, emphasis on implementation over user-centric design can lead to design fixation ~\cite{alipour2017review}; users may be led down one path of technical implementation, but lack the divergent exploration that can enable them to find the most fitting approach 
%and interaction paradigm 
for their problem ~\cite{CSDURAO2018265}. 
These applications may function well technically, but 
% The emphasis on implementation over user-centric design leads to solutions that may function well technically, but these solutions 
can lack the depth of refinement that helps ground a prototype in real user needs.
% needed to create engaging and intuictive user experiences.
%Tools focused on implementation do not support for true exploratory programming and %\vivian{lack a conceptual zooming out of the code level that can help ground a prototype in real user needs}. 

In exploratory programming, it is crucial to explore different solutions
%ways of addressing a problem 
before implementation~\cite{cstschneiderman}. Our approach is to integrate design with code synthesis. Before implementing a solution, we consider many factors, such as 
the user (who are they and why do they want this?), the core approach (for a learning application, are you following a learning theory like spaced repetition or generation and elaboration?), and the interaction paradigm (if you are building a crowd-sourced hotel search platform, are you basing it off of an existing paradigm like Uber, or a card-swipe paradigm like Tinder for low cognitive load?). 
%These are all important to consider and in order to evaluate them, users must prototype it with a real application and user data. 
By providing a structured framework for design exploration, we can guide users to collaborate with LLMs in exploring a problem space. Combined with their code-generation capabilities, LLMs can assist users in creating user-centric, code-based UIs that serve as both functional prototypes and stepping stones towards final products. 
% space. Our work places the end-user at the forefront of the design process, requiring empathy and understanding of user challenges, and iterative refinement based on real-world feedback.

%Additionally, exploring a design space requires both divergent and convergent thinking ~\cite{muller2011leaving}. 
%Divergent thinking prevents design fixation by encouraging users to consider multiple options, while convergent thinking is necessary to synthesize and evaluate concepts to anchor them in practical applications. 
%LLMs have the capacity to assist with this. They can almost instantly produce hundreds of outputs, inspiring users to broaden their ideas ~\cite{10.1145/3490099.3511105, kargaran2024menucraftinteractivemenudesign, 10.1145/3544548.3580907}. 
%LLMs can also assist users in refining concepts
% producing polished and detailed outputs from vague ideas 
%by crystallizing abstract problems into concrete design solutions. 
%Combined with their code-generation capabilities, LLMs can guide users in creating user-centric, code-based UIs that serve as both functional prototypes and stepping stones towards final products. 
%\vivian{I think this paragraph and the one before it have to combine somehow. I would make the diverge/ converge a little secondary to the angle about PAI. I did a double take at what seemed like another theory/ motivation paragraph }

%DESIGN, CODE, DIVERGENT CONVERGENT THINKIN

We present DynEx, an LLM-based method for exploratory programming for functional UIs.
%rapidly exploring UIs.
%from initial ideas of varying levels of specificity \vivian{better to be clear. about the levels of specificy. it's two levels right?  }.
DynEx helps the user build code-based UI prototypes in two stages: (1) structured design exploration through a Design Matrix, and (2) \revision{LLM code generation through Modular Stepwise Implementation}.
The Design Matrix brainstorms unique ideas across three dimensions (Person, Approach, and Interaction) to help users explore the design space. The idea is then grounded, a process to synthesize abstract ideas into detailed application designs. The system next breaks down the project into \revision{modular} steps to execute sequentially. \revision{Each step is distinct and nonoverlapping, allowing users to modify and test code at each step without worrying about unintended changes affecting other steps.} Our prototypes self-invokes multi-modal LLMs to produce applications that have larger sets of synthetic placeholder data, can generate images, and provide recommendations, enabling the creation of a diverse suite of applications.
%1)The first stage utilizes LLMs to help explore a design space through a user-centric structured framework. It’s structured in the form of a design matrix that guides the user to think through (1) the person the application is for, (2) the approach or methodology driving the solution, and (3) the interaction paradigm or user experience of the application. It does so in two ways: (1) divergent thinking through \textit{idea} generation, where the system brainstorms multiple unique ideas to help users explore the design space, and (2) convergent thinking through \textit{grounding}, a process that synthesizes abstract ideas into application designs anchored in human-centered design. 
%The second stage leverages LLMs to dynamically and iteratively build code-based UI prototypes. From the Design Matrix, an implementation plan is created and executed iteratively through multiple steps. The system supports the creation of simulated data and self-invokes LLMs so that the generated code can call GPT and DALL-E 2, enabling the creation of a diverse suite of applications, such as: language learning app for busy parents, a music taste exploration app for niche-genres, a calorie and macros tracker, and a git command quiz generator. Users iteratively generate and refine the code, and can transition between various working states until the prototype is complete. 

Our contributions are as follows:
\begin{itemize}
\item DynEx, an LLM system for accelerated exploratory programming where the user inputs a problem they want to solve and rapidly creates UI prototypes. The system guides users through a design space, then allows users to sequentially generate and interact with UI code.
Users can create multiple variations of prototypes and iteratively refine their ideas. 

\item The Design Matrix, a structured framework using LLMs to guide users through the design space by considering the Person, Approach, and Interaction through idea generation and grounding.

\item Modular Stepwise Implementation, a simple technique for building code-based UI prototypes where the implementation plan is broken down into steps and code is generated sequentially. At each step, users can edit or approve the code before moving onto the next step. Each step is modular, meaning that it is distinct, nonoverlapping, and testable.
%Users can users to fallback to previous steps. 
\item A qualitative evaluation of 10 experts, demonstrating that DynEx enabled them to explore the design space and create more complex and varied prototypes compared to a Claude Artifact baseline where users must lead the design process themselves.
\end{itemize}
\section{Related Works}

\subsection{Exploratory Programming and Prototyping}

Exploratory programming is a practice in which programmers actively experiment with different possibilities using code ~\cite{8103446}. It is grounded by five main characteristics: (1) Needs for Exploration, (2) Code Quality Tradeoffs, (3) Ease or Difficulty of Exploration, (4) Exploration Process, and (5) Group or Individual Exploration. Like other prototyping methods, it is important to consider the cost of exploratory programming compared to its value in providing insights ~\cite{tiong2019economies}. Exploratory programming for UI creation must be easy and quick but still produce functional code. It must also produce sufficiently complex interfaces that can actually inform designers on how users will interact with them. Finally, it must support lightweight version control so that individuals can easily see the variations they create ~\cite{kery2017variolite}.

Prototyping and exploratory programming are forms of creative problem-solving. During exploratory programming, designers iteratively test and refine ideas via code. Generating multiple alternatives, exploring directions, and reverting to previous versions have been labeled as requirements for creativity support tools~\cite{cstschneiderman}. Past research demonstrates the effectiveness of prototyping during design~\cite{parallelproto, CSDURAO2018265}. Learning theory research suggests that generating variations of the output enables critical reflection and deepens understanding in problem domains~\cite{marton1997learning}. Prototyping has been shown to be an effective means of inquiry to increase the granularity of design requirements based on early user feedback ~\cite{wensveen2014prototypes, KANG2023101147}. However, it is not always possible to prototype UIs in low-fidelity mediums. Applications often change requirements and need to support complex user interactions ~\cite{10.1145/3613904.3642774}. Even higher-fidelity approaches like Figma ~\cite{figma} have limitations in modeling the interactive experience of a UI. It can lack the richness of user experience that comes from simple backend computation or data processing ~\cite{8103446} and have other time and resource costs ~\cite{tiong2019economies}. Exploratory programming addresses these issues, allowing for the prototyping of code-based interactive applications.

\subsection{\revision{Design Space Exploration with AI}}

\revision{
HCI literature describes four classic classes of design space exploration \cite{shireen2020bridging,davis2024fashioning}. 
First, \textit{example-based exploration} helps users ideate solutions by searching a database of related examples to inspire them  \cite{chou2023talestream, koch2019may, kang2021metamap} or help find flaws in their ideas or applications \cite{10.1145/3613904.3642335}.
Second, \textit{reflective exploration} provides a rich history-keeping mechanism that facilitates reflection during ideation when needed \cite{klemmer2002web}.
Third, \textit{genetic exploration} generates new solutions by combining components of existing examples \cite{kim2022mixplorer}. 
Finally, \textit{dimensional exploration} provides a mechanism for breaking down a design space into fundamental dimensions. It first explores multiple ideas for each dimension separately, then synthesizes ideas from each dimension to form a cohesive solution~\cite{micheli1997solution, walker1997solution}.

% Using LLMs for ideation in creativity tools has been explored in music ~\cite{10.1145/3313831.3376739}, visual design ~\cite{liu2022opalmultimodalimagegeneration, wang2023popblendsstrategiesconceptualblending}, game design \cite{Lomas2021}, and writing ~\cite{10.1145/3491102.3501819, 10.1145/3586183.3606800, 10.1145/3544548.3580907, gero2021sparksinspirationsciencewriting, 10.1145/3462204.3481771}. 

%LLM-based creativity support tools must have sufficient ability for users to create and compare numerous variations along different dimensions of the design space \cite{cstschneiderman}. 
%Dimensional exploration is particularly helpful in the initial phases of design exploration in enabling designers to break down complex problems into key aspects, evaluate trade-offs, and explore different design directions and view the problem at a higher, conceptual level ~\cite{Beaudouin-Lafon2007, Heape2007}. 
%An LLM-system based on dimensional exploration can explicitly state (and allow a user to dictate) which dimensions were prioritized, how trade-offs were made, and what logic was used to generate specific UI components. 
Dimensional design exploration has been investigated across various systems. 
It poses a significant challenge for individuals attempting it independently ~\cite{10.1145/3059454.3059472}. 
When left to their own devices, people tend to prematurely converge on ideas without fully considering all possibilities ~\cite{CROSS2004427, parallelproto, janis1982groupthink, JANSSON19913}.
Existing systems use LLMs to help with dimensional design space exploration. 
For example, Spellburst~\cite{angert2023spellburst} helps users navigate a multi-dimensional space of generative art possibilities for refining the current design.
Luminate~\cite{suh2024luminate} facilitates the initial exploration of narrative design spaces for users by generating dimensions with LLMs; however, the dimensions are not controlled and sometimes result in incoherent outputs.
We focus on the dimensional approach for initial design space exploration through DynEx's Design Matrix, which allows users to thoroughly think through their problem in three fixed dimensions (Person, Approach, Interaction) before implementation, which is showcased in Section 3.
The already-defined dimensions in the Design Matrix are used as context to brainstorm relevant ideas for the current dimension, so that the separate dimensions synthesize into a coherent prototype design. 

}

\subsection{LLM-Based Tools for Code Generation}
\subsubsection{LLM-Based Code Synthesis Tools}

State of the art models for code generation include GPT-4, AlphaCode, CodeGEN, Code Llama, and Gemini ~\cite{openai2023gpt4, li2022alphacode, nijkamp2022codegen, code_llama_meta, google_gemini}. These models generally perform well in transforming a natural language problem specification into a simple code solution. They are most effective at creating code segments for specific functions or features. Systems built around these LLMs show promise for improving LLM-code generation beyond direct prompting. However, many of the UI code generation tools have limitations. For example, Co-Pilot ~\cite{github_copilot} is a system that can modify existing code repositories, but it does not support end-to-end code generation. OpenHands ~\cite{wang2024opendevin} and GPTPilot ~\cite{gpt_pilot} are open-source agent-driven systems for end-to-end software development. They use separate LLM agents to represent various roles within the software development process and implement a technical specification. 
%\sout{However, these systems lack the support for exploratory programming that is embedded in DynEx, which helps users explore during the planning and problem specification stages of development.}

\subsubsection{LLM-Based Code Synthesis Tools for UIs}

LLMs have been proven to be especially good at creating UI code.
Specifically, LLMs have been shown to be extremely apt at writing code from provided UI screenshots and designs ~\cite{lu2023uilayoutgenerationllms, soselia2023learninguitocodereversegenerator, si2024design2codefarautomatingfrontend, xiao2024_prototype2code, qian-etal-2024-visual, wu2024uicoderfinetuninglargelanguage, wan2024automaticallygeneratinguicode, yun2024web2codelargescalewebpagetocodedataset} and natural language prompts; WebSim is a system which enables users to created simulated UIs with minimal prompting ~\cite{websimai}. However, these systems do not offer users much design support, placing the brunt of ideation and design space exploration on the user. DynaVis is a tool that dynamically synthesizes widgets for data visualization UI. While it allows users to make dynamic components, it does not support the creation of a complete user interface~\cite{vaithilingam2024dynavisdynamicallysynthesizedui}. 

Claude Artifact ~\cite{claude_3_5_sonnet} likely offers the best out-of-the-box solution for UI prototyping. It creates functional UIs from conversational prompts, provides users with a window for rendering, and keeps track of version history. However, Claude Artifact does not guide users through the design exploration process. The onus is on the user to use repeated prompting to add new features and move towards a solution. Given that LLMs have demonstrated the ability to effectively power creativity support tools and perform divergent design space exploration in other domains ~\cite{10.1145/3613904.3642400, 10.1145/3490099.3511105, 10.1145/3544548.3580907}, it should be possible to build LLM systems that support both design exploration and implementation for UI prototyping.

% However, systems for this task do not offer any design support. They are specifically for reconstructing already existing UI designs or screenshots in code. DynaVis is a system that allows users to dynamically synthesize widgets for visualization editing, creating user interface components through natural language, but does not allow users to create complete UIs, only specific components ~\cite{vaithilingam2024dynavisdynamicallysynthesizedui}. These systems highlight the lack of existing systems which incorporate design exploration into end-to-end UI code generation.

% \todo{LLMs are good at X. They are especially good at creating from scrrenshots. And widgets. There is potential to do what we want which is explore the design space. THOUGHT: a lot of these systems take in screenshots, but don't take into account design. DynaVis doesn't fit here. Throway sentence like: "systems like dynavis do dynamically synthesize widgets, but don't do full UIs" -theres potential to explore a design space and do what we want wwith it ADD CODE GENEARITON}. 

\subsubsection{\revision{Tool-Calling with AI}}
\revision{Tool-calling has been demonstrated to improve the outputs of LLM-based systems. It refers to the ability to interact with external tools, APIs, plugins, or even other LLMs to enhance their functionality beyond text-based responses \cite{yang2023gpt4tools}. The difficulty in sourcing example data has limited the success in prototyping with LLMs ~\cite{10.1145/3491101.3503564}. Tool-calling is integral in making LLMs more interactive and useful in complex, real-world applications}. WebSim calls LLMs to generate images for created UIs ~\cite{websimai}. Jigsaw is a system for prototyping that uses puzzle pieces as metaphors to represent AI foundational models ~\cite{lin2024jigsaw}, allowing users to chain foundational model capabilities for the creation of more complex outputs. 
\revision{We use tool-calling in DynEx by enabling our prototypes to self-invoke generative AI, allowing for the creation of images, text, or data, and emulate recommendation or sensemaking systems.} 
%DynEx therefore has specific functionality to function-call LLMs to easily create data for applications.

\subsubsection{\revision{Task-Breakdown and Structured Prompting with AI}}
%There are several other factors we consider more broadly that are relevant to LLM-systems. First, m
Many tasks are too complex for LLMs to accomplish in one prompt and have to be broken down into smaller problems ~\cite{wu2022promptchainerchaininglargelanguage, DBLP:journals/corr/abs-2110-01691}. LLMs are also more successful in accomplishing tasks when asked to create a plan and reason through intermediate steps ~\cite{cai2024lowcodellmgraphicaluser, kojima2023large, suzgun2022challenging, DBLP:journals/corr/abs-2201-11903}. Creating a complete application is a involved task; DynEx decomposes the project into steps to create an implementation plan that can be iteratively executed. 
\section{\revision{System Overview}}
We introduce DynEx, an LLM-based system for accelerated exploratory programming that assists users with design exploration and code implementation.
The input is the problem that the user wants to solve. It can be vague, such as an application to help with plant-watering or meal-planning.
%; something that the user does not have a clear solution for. 
\revision{The output is a functional React application that addresses the user's problem. It integrates generative AI to create images and data that the application requires, enabling the creation of a realistic prototype suitable for user interaction. The code is contained within a single file HTML page written in Javascript, HTML, and CSS, and is limited to 450 lines of code due to the 4096 token constraint from Claude 3.5 Sonnet, which we used for code generation. }
%, that also utilizes React ~\cite{react} and Material UI Library ~\cite{materialui}.
%\revision{The code output is contained within a single file and limited to 450 lines. }

DynEx is designed for programmers who have personal projects they want to explore. Many of these ideas remain underdeveloped or unrealized for years, if not indefinitely. Bringing these ideas to fruition is challenging because it requires not only developer experience to prototype the idea in code, but also addressing gaps in the design. DynEx helps concretize users' ideas through design exploration and implementation, allowing them to take a crucial step towards creating a minimal viable product (MVP).

The system has 2 stages: 
\begin{enumerate}
    \item \textbf{Design Exploration through the Design Matrix}, where the system helps the user identify the person, approach, and interaction paradigm of the application.
    %, the details of which are defined below.
    %using idea generation and grounding, a method of specifying ideas to become a more solidified design.
    \item \revision{\textbf{Code Generation through Modular Stepwise Implementation}, where the system creates a multi-step plan with modular steps, and code is generated for each step sequentially. The user tests and approves the code at each step before the system generates code for the next step.}
    %\todo{ and the system generates code for the first step, the human tests and modifies code for this step before approving it. Then the system generates the code for the next step, and the human tests, modifies, and approves, etc. }}
    % until the creation of the final prototype. The user can test, approve, and modify the code at each step.}
    %Our implementation process supports the generation of synthetic placeholder data and 
    %The prototype self-invokes multi-modal LLMs, a technique for the generated code to call other LLMs to create a dynamic and diverse suite of applications.
\end{enumerate}

%A user begins by suggesting a problem they would like to solve, such as “I need reminders water my plants” or “I need to plan my meals”-- it can be vague, something they have not thought of a solution for. 
\revision{This remainder of this section introduces the main concepts behind the Design Matrix and Modular Stepwise Implementation. Section 4 then illustrates a usage scenario featuring a user developing a prototype with DynEx, followed by Section 5, which discusses DynEx's implementation details.} 

\subsection{Design Exploration through the Design Matrix}

\begin{figure*}[h]
    \centering
    \includegraphics[width=\textwidth]{figures/system_matrix_diagram.png}
    \caption{\textbf{The Design Matrix:} There are three columns for the Person, Approach, and Interaction dimensions. There are two rows for the levels of specificity: Idea and Grounding. The matrix can be traversed in any order as long as the Idea comes before the Grounding for each dimension. Each entry will take all already-submitted entries as context to generate a response. For example, the \textit{Interaction:Grounding} will take in all previous entries as context in this diagram. }
    \Description{A diagram of DynEx's Design Matrix. There are three columns for the Person, Approach, and Interaction dimensions. There are two rows for the levels of specificity: Idea and Grounding. The matrix can be traversed in any order as long as the Idea comes before the Grounding for each dimension. Each entry will take all already-submitted entries as context to generate a response. For example, the \textit{Interaction:Grounding} will take in all previous entries as context in this diagram. 
    }
    \label{fig:system_matrix_diagram}
\end{figure*}
%Once a problem submitted, the user is led through the Design Matrix, a structured framework that guides users through the design space. 
The Design Matrix is a structured framework that facilitates the transformation of abstract concepts into actionable design components.
% allowing the user to understand their problem and solution space before settling on a prototype design
\revision{It is a crucial part of exploratory programming, allowing users to explore their problem and solution space
%, map out design possibilities,
and balance trade-offs between design choices before settling on a prototype design. The Design Matrix is rooted in the dimensional approach to design space exploration. The matrix organizes complex design challenges into three distinct dimensions to ensure that all these facets are thoroughly considered and addressed before application development. Three key design dimensions make up the columns:}
\begin{enumerate}
    \item Person - who the application is for. 
    \item Approach - the method, theory, or strategy behind the solution. 
    \item Interaction - \revision{the interaction paradigm that the user interface is implemented around. For example, a card-swipe, news feed, or table layout.}
\end{enumerate}
Each dimension is explored on two levels of specificity, which make up the rows: 
\begin{enumerate}
    \item Idea - \revision{a phrase that conveys the main concept of a dimension. For example, in the Interaction dimension, one idea could be "flashcards implemented in a card-swipe UI."}
    % the brainstorming of the higher-level concepts across each dimension. 
    % generates concise and diverse options across each dimension to give users a quick view of the potential solutions
    \item Grounding - \revision{3-5 bullet points that give concrete details for how the idea is realized.
    %the solidification of foundational details that specifies how the idea should be developed. 
    For example, the grounding for the card-swipe idea should specify what information is displayed on each card, what action is associated with swiping right, and what action is associated with swiping left. }
\end{enumerate}
%Fig.~\ref{fig:matrix} illustrates the Design Matrix, with columns for Person, Approach, and Interaction, and rows for Idea and Grounding. 

\revision{

The \textbf{Person} dimension describes who the user is, what needs they have, and why they are creating this application. In human-centric design, understanding a person's context and motivations is key to creating an application truly tailored to user needs. For example, potential users of a language learning application might range from a business traveler to a student interested in cultural immersion. A business traveler prioritizes practical, situational language skills tailored to business communication. In contrast, a student interested in cultural immersion prioritizes conversational fluency and deeper engagement with aspects such as history, pop culture, and cuisine.
%when learning a new language, there can be many potential users, from a beginner, intermediate, or advanced language learner to someone interested in cultural immersion. Figuring out the user's motive is imperative to inform later stages of the design. The Person dimension that explores the user's goals and challenges within the broader context of the problem they want to solve.

The \textbf{Approach} dimension describes the underlying mechanism to achieve the user's goal; it details the theory, algorithm, or model that is the essence of the solution to the problem. For example, in a learning application, learning theories that can drive an application include spaced repetition, a learning technique to review information at gradually increasing intervals to enhance long-term memory retention, and gamification, a technique that uses game elements to encourage learning. 

The \textbf{Interaction} dimension describes how information should be displayed and engaged with in the UI. 
%Finally, interaction paradigms offered in UIs all have different affordances and tradeoffs, and thus DynEx allows user to explore them through the \textbf{Interaction} dimension of the matrix. 
For example, for a flashcard application, a card-swipe interaction complements a spaced repetition approach because it provides a natural, dynamic way for users to move through cards at their own pace. Other potential interactions that pair with spaced repetition could include a calendar-based interface to schedule learning reminders, or even a news-feed interface with quizzes.
%The Interaction dimension allows users to explore different interaction paradigms and evaluate different UI concepts. While we could have chosen other dimensions, we found the Person, Approach, and Interaction, to be most central to application development.

In the Design Matrix, each dimension has two levels of fidelity: \textbf{Idea} and \textbf{Grounding}.
%so that we can explore them on two levels of fidelity: concept and realization.
%, we explore them on two levels of granularity, which is showcased in the two rows of the the Design Matrix: Idea and Grounding. 
The idea is a low-fidelity concept that is simple to understand, but lacks precision. 
%The goal is to provide users with a brief overview of different design perspectives and quickly identify promising concepts. 
The grounding is a higher level of fidelity that expands the idea, providing details that specify how to fully realize this idea in a prototype. It paints a more comprehensive vision of the design.
%Both levels are essential for users to explore the problem space and gain a comprehensive vision of the design.
%In the Idea row, the Design Matrix generates concise and diverse options across each dimension to give users a quick view of the potential solutions. 
%While these generated ideas are short and easy to read, they are insufficient to translate into code. The Grounding row provides a more specific and precise version of the idea the user select's. It fully expands an idea, adding more context and details to paint a more comprehensive vision of the design. 

%DynEx uses context curation to guide the outputs of each entry in the matrix. 
For each entry in the matrix, LLMs generate suggestions the user can accept, edit, or regenerate. The Design Matrix uses all its previously-submitted entries to make suggestions for the current entry. This happens both across rows, where the grounding depends on the idea, and across columns. This ensures that the current design direction is considered when brainstorming suggestions for each entry and that ideation across dimensions is both consistent and relevant, ultimately leading to a cohesive design solution. For example, if the Approach (ie: spaced repetition) is already filled out, the system will suggest a corresponding Interaction (ie: flashcard swipe).
Together, these 6 entries describe the prototype design in enough detail that it can be implemented in code.
%it uses information from all the other cells that are filled in to inform it. WHY -- it's because it needs to be cohesive. This happens both across rows (the grounding depends on the selected idea) and across columns. The user can traverse the columns in any order, and the already-submitted columns will influence the current column.  There are gaps between ideas and interactions -- if the person says they are a visual learner than the interaction should suggest consistent ideas like flashcards. CONSISTENT COHERENT RELEVANT}

%Generating outputs for an entry depends on other entries in the Design Matrix. 
%Idea entry to know what to expand. The user can traverse the columns in any order, and Across columns, the matrix will use the already-specified dimensions Thus, when it suggests interaction paradigms, it is aware of the approach and suggests appropriate interaction paradigms.
}

\subsection{\revision{Code-Generation through Modular Stepwise Implementation}}
%\revision{Once the Design Matrix has been traversed, DynEx leverages LLM code-generation capabilities to create the prototype.} The system first identifies the project requirements, creates the project specification, and generates synthetic placeholder data if required. Once the initial setup is complete, the user and system collaborate in creating the prototype through \revision{Modular Stepwise Implementation}, where the system breaks down the implementation plan into steps. 
%(see Fig ~\ref{fig:system_implementation}- F).

\subsubsection{Project Scoping}
\revision{Before implementing the prototype, DynEx must (1) identify technical requirements and (2) generate a project specification (spec). DynEx uses the contents of the Design Matrix to identify technical requirements. If the prototype needs to support data visualization and flow charts, external libraries, such as ChartJS and GoJS, can be used. If the prototype needs data generated on-the-fly, such as for personal recommendations, the prototype will call GPT. If images are required, the prototype will call DALL-E 2. Identifying technical requirements prior to development minimizes ambiguity and missteps during implementation.
%Will the prototype need to call external libraries like ChartJS or GoJS to support data visualization and flow charts? Will the prototype need to call GPT to support personalized recommendations? Will it need to call DALL-E 2 to generate images? It's necessary to define these technical prerequisites before development.
}

%(see Fig ~\ref{fig:system_implementation}- C, D E).
%The system brainstorms these requirements using the Design Matrix as context. 
%Users can manually update and change the project requirements if needed - for example, if the application is a book recommendation app, and DynEx did not recommend the images as a requirement, the user can add that requirement themselves if they so please. 

\revision{DynEx uses the technical requirements and the Design Matrix to generate a detailed, bullet-point spec, typically about one page long. Specifically, the spec details the application layout (what content will preside in the interface), the user interaction (how the application will respond when a user clicks on buttons), and logic of the application (when and how the application will call LLMs or use external libraries). It outlines the exact functionalities of the application. 
%If the project requirements dictate that data is needed, the spec helps define the schema. If project requirements dictate that GPT must be self-invoked, it details what the code should prompt for and the expected response. 
%Ultimately, the spec transforms the project requirements and contents of the Design Matrix into actionable implementation instructions, ensuring the functionality of the app is well-aligned with the user's goals; it provides a clear roadmap for development.
}
% Although there is potential overlap between the spec and Grounding level in the Design Matrix, the project specification focuses more on the technical requirements for the project. 
%DynEx breaks the spec down into steps to iteratively implement on later on in the workflow. 
%\paragraph{Synthetic Placeholder Data}
%We leverage few-shot examples with LLMs to guide the generation of relevant and appropriate placeholder data. 
%The user can regenerate, approve, or modify the outputs. 

\subsubsection{\revision{Modular Stepwise Implementation}}
\begin{figure*}[h]
    \centering
    \includegraphics[width=\textwidth]{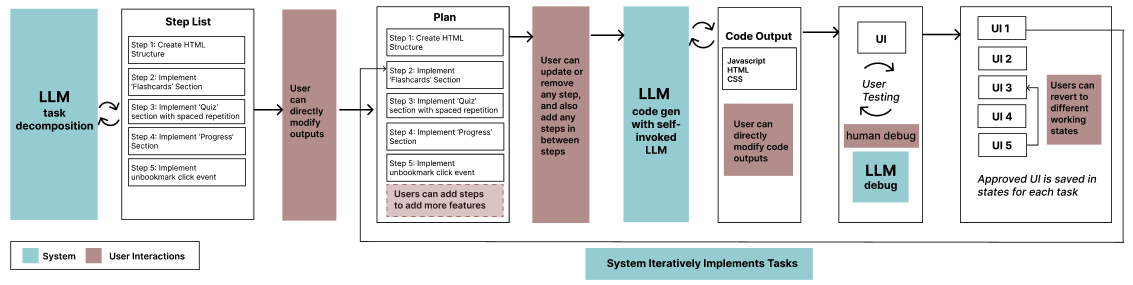}
    \caption{\textbf{DynEx Code Generation:} The system first decomposes the project into steps that are modifiable by the user. The user iteratively implements each step. The user can add, remove, or update steps at any point. The backend LLM then generates the UI code -- the user can also interact with the UI . The user can directly modify the code output, debug, and revert to previous versions. The user iterates through steps until the final prototype is complete.}
    \Description{A diagram of DynEx's code generation process. The system first decomposes the project into steps that modifiable by the user. The user iteratively implements each step. The user can add, remove, or update steps at any point. The back-end LLM then generates the UI code -- the user can also interact with the UI . The user can directly modify the code output, debug, and revert to previous versions. The user iterates through steps until the final prototype is complete
    }
    \label{fig:system_implementation_diagram}
\end{figure*}

To implement the prototype, DynEx employs a \revision{modularized, stepwise} approach to development. \revision{It breaks down the spec into a multi-step plan, where each step is the next-smallest testable unit of the previous step.}  
% \revision{\todo{
% Each step is nonoverlapping and distinct, and builds incrementally on what has already been completed. This modularity allows users to regenerate and modify the code at any step without worrying if leftover changes will affect previous steps. }}
At each step, users test and approve the code before moving on to the next until the completion of the prototype. 
%Each step is executed sequentially, and the user tests and approves each step before moving on to the next. 
%The user can regenerate or directly modify the code to help it recover from failure at each step. 
They can also edit, remove, and add steps at any point in the implementation process.

%The system breaks down the spec into steps, where each step is the next-smallest testable iteration of the previous step.
DynEx generates a multi-step plan because specs are complex and detailed, and attempting to generate code with a one-shot approach often results in key elements being missed. 
%With missing functionalities, users can spend unnecessary time debugging and adding features to align with the initial spec. 
%Breaking down the spec into manageable steps ensures that details are not missing in the final prototype. 
\revision{
Each step is modular and nonoverlapping, similar to how developers incrementally adds features that build on prior progress. The first step establishes the basic HTML structure. The following steps add functionality in a logical order, allowing the system to generate code iteratively without needing to rework or disrupt the code before it. For example, when creating a book recommendation application, the process starts by setting up the HTML structure, building different sections of the application, and finally tool-calling GPT. 
}

%Furthermore, our iterative step-by-step approach saves the code at each stage, providing a natural form of version control.
% \paragraph{Feature Adding and Version Control}
%We allow users to update, remove, and add steps to our step list as DynEx generates code for each step. 

The modular steps also provide version control, a necessary functionality in exploratory programming systems.
Since each step is saved, DynEx provides a clear version history that captures the most advanced working prototype while offering the flexibility to iterate between versions. 
%This allows for structure and safety in an unpredictable process, and protects users from irreversible errors.
%Users can easily revert to prior working states if they want to reconsider a feature, without worrying if leftover changes will affect the process.
%, creating a simplified form of version control.
%While this is not a main contribution of our system, 
%Since LLMs can sometimes produce faulty or unusable code, this stepwise approach serves as a safety net where u
Users can always fall back on the last successful version, ensuring that they always have a stable point to return to. 

\subsubsection{Self-invoked Multi-Modal LLMs} 
\revision{Tool-calling is a known method to make LLMs more powerful by having them interact with external APIs. In DynEx, the generated code can call the OpenAI API to generate data and images as needed. For example, a book recommendation system needs to call an LLM after a user has selected preferences to get recommendations. An outfit recommender system will call image models to generate previews of the user's outfits based on their existing wardrobe. We call this method self-invoking multi-modal LLMs. It enables the creation of powerful and realistic applications that can generate data to respond to user intent, a capability not offered by prototypes supported by static data.}

\subsubsection{Creating Multiple Prototypes}
Once a prototype is created, users can create additional variations by revisiting the Design Matrix, tweaking parameters to explore variations, and comparing the results. 
Users can modify entries in the Design Matrix and re-implement new designs to generate a diverse set of prototypes. 
This is a consistent method with exploratory programming, empowering users to explore multiple solutions in parallel and refine their designs. 
\section{\revision{Usage Scenario}}
%A user begins by suggesting a problem they would like to solve, such as “I need reminders water my plants” or “I need to plan my meals”-- it can be vague, something they have not thought of a solution for. 
\revision{To illustrate the user experience of DynEx, we present the example of a user, Franky. Franky wants to create a personalized application to learn Chinese that addresses his specific learning needs, but does not have a clear design in mind. He starts by typing "learn Chinese" in the "Problem" box, and clicks "Submit" (Fig \ref{fig:matrix} - A). }
%We use a motivating example to illustrate the process of creating a prototype using DynEx. \revision{The user wants to create an application to learn Chinese, yet does not have a clear design in mind.} The user inputs the problem in the input box (): to learn Chinese.

\subsection{Design Exploration through the Design Matrix}

\begin{figure*}[h]
    \centering
    \includegraphics[width=\textwidth]{figures/system_matrix.png}
    \caption{\textbf{DynEx's Design Matrix User Interface}: Users traverse through the matrix, which guides them through exploring the Person, Approach, and Interaction dimensions relevant to the problem space. For each dimension, users begin on the Idea level by inputting their problem (A). They then brainstorm ideas (B), and can iterate on the ideas (C). They can select an idea (\revision{D}) and submit it in the input box (E). They then move on to the Grounding level, brainstorm a response (F), and submit that as well. They have the option to save versions (G), in case they want to move back to the Idea level and explore a new idea. Previous entries used to curate the current entry response are highlighted in yellow \revision{(I)}. Finally, users can can explore the prototype (H). }
    \Description{Screenshot of DynEx's Design Matrix user interface annotated with captions on each side of the interface that call out important features. Users traverse through the matrix, which guides them through exploring the Person, Approach, and Interaction dimensions relevant to the problem space. For each dimension, users begin on the Idea level by inputting their problem (A). They then brainstorm ideas (B), and can iterate on the ideas (C). They then select an idea (C) and submit it in the input box (E). They then move on to the Grounding level, brainstorm (F), and submit that as well. They have the option to save versions (G), in case they want to move back to the Idea level and explore a new idea. Previous entries used to curate the current context are highlighted in yellow. Finally, users can can explore the prototype (H). }
    \label{fig:matrix}
\end{figure*}
\revision{The Design Matrix appears on the screen. He sees three columns for the Person, Approach, and Interaction dimensions, and two rows for Idea and Grounding -- the two levels of specificity. To reference entries in the matrix, we use a \textit{column:row} notation; for example -- \textit{Person:Idea}, \textit{Approach:Grounding}.}

%The user is greeted with the Design Matrix. 

\subsubsection{Person}
\revision{To choose an the idea for the Person (\textit{Person:Idea}), Franky clicks the "Brainstorm" button(Fig \ref{fig:matrix} - B). After a moment, three potential target users appear in that entry (Fig \ref{fig:matrix} - D). }
\newline\newline
\noindent
\noindent
\begin{tabular}{c c c} 
    \fbox{
        \parbox[c][2cm][c]{0.12\textwidth}{
            \centering 1. Non-native speakers interested in Chinese culture
        }
    } & \setlength{\fboxrule}{2pt}
    \fbox{
        \parbox[c][2cm][c]{0.12\textwidth}{
            \centering 2. Visual learners struggling with language memorization
        }
    } & \setlength{\fboxrule}{0.5pt} 
    \fbox{
        \parbox[c][2cm][c]{0.12\textwidth}{
            \centering 3. Travel enthusiasts planning a trip to China
        }
    }
\end{tabular}\\~\\
\revision{Franky can generate as many ideas as he wants \revision{(Fig \ref{fig:matrix} - B)} or iterate on them and re-brainstorm \revision{(Fig \ref{fig:matrix} - C)}. Curious to see more suggestions in the Person dimension, Franky clicks the "Brainstorm" button again, and 3 more ideas are added:} 
\newline\newline
\noindent
\noindent
\begin{tabular}{c c c}
    \fbox{
        \parbox[c][2cm][c]{0.12\textwidth}{
            \centering 4. Retired adult wanting to expand linguistic skills
        }
    } &
    \fbox{
        \parbox[c][2cm][c]{0.12\textwidth}{
            \centering 5. Busy university student wanting to study Chinese in his free time
        }
    } & 
    \fbox{
        \parbox[c][2cm][c]{0.12\textwidth}{
            \centering 6. Chinese-born American looking to brush up language skills
        }
    }
\end{tabular}\\\\
% The user collaborates with the system to brainstorm who the application is for in \textit{Person:Idea}. The system brainstorms potential target users and returns 3 results:
\revision{Franky selects (2) on the interface -- he identifies as a visual learner, recognizing that many of his learning challenges stem from lack of visual aids. \textit{"Visual learners struggling with language memorization"} shows up in the input box (Fig \ref{fig:matrix} - E), which he can directly update if he wants.}
%Each of these cases have nuances; for (1) the user would probably be more interested in creating an app to learn Chinese culture rather than Chinese words; for (2), the user prefers a visual learning application; for (3), the user would probably focus more on learning travel-related vocabulary and phrases. 
% For example, if the user wanted more results that focused on learning Chinese culture, they could type that in the iteration box and brainstorm ideas along that direction. 
% All of these results have different implications as well. For (4), the retired adult presumably has more time to learn Chinese -- their method for learning would be different than (5), the busy university student who can only study at short intervals. Additionally, (6) presents a different use-case, a user who already has existing Chinese knowledge. 
%\revision{The user brainstorms more ideas (Fig \ref{fig:matrix} - D) -- all the idea generations can be seen in Appendix \ref{app:usage_matrix}.}  Ultimately, the user selects (2) \textit{"Visual learners struggling to learn with language memorization"}.
\revision{He accepts the idea as is, excited to see more details about this in the grounding row (\textit{Person:Grounding}). }
%Franky clicks the "Brainstorm" button in that matrix entry (Fig \ref{fig:matrix} - F). The idea entry (\textit{Person:Idea}) turns yellow on the interface, since the system uses it as context to generate the grounding \revision{(Fig \ref{fig:matrix} - I)}.}
%to iteratively deepen their idea. 
\revision{A list of bullet points that motivate why a visual learner might want this application appear in the grounding entry on the interface}: 
\begin{center}
\fbox{
    \parbox{.9\linewidth}{
        \begin{itemize}
            \item Confusion arises due to unfamiliarity with Chinese characters and their complex structure, slowing the learning process.
            \item Difficulty in linking characters to their corresponding meaning or pronunciation, hindering vocabulary acquisition.
            \item Traditional memorization methods offer little aid to visual learners who could better recall information through imagery.
        \end{itemize}
    }
}
\end{center}
\revision{This reasoning resonates with Franky, especially the second bulletpoint. As a visual learner, his struggle with traditional text-based language learning stems from the lack of meaningful visual connections that aid memory. Franky has the option to directly edit the grounding, save this version, or iterate on it, but he accepts the grounding as is, excited to build a learning application with a visual approach. He moves on to explore the Approach column in the matrix.}
%When grounding the Person dimension, the system aims to fully understand the user goals, challenges, and the broader context in which the problem should operate. 
%\textit{Person:Grounding} also defines the specific problems that users face, some shortcomings of existing solutions, and how the application would address these gaps.  
%The user can directly edit, save this version, or iterate on it. \revision{The user is satisfied with this grounding,} so they submit their result and move on to the Approach dimension. 

\subsubsection{Approach}
%and the system brainstorms methods or strategies that the application could employ to solve the problem, 
\revision{Franky generates ideas for the Approach dimension (\textit{Approach:Idea})
%, and 
%and sees that the entries for the already-submitted Person dimension have turned yellow, used as context to generate Approach ideas. 
%three Approach ideas that support visual learning appear
:}
%\textit{Person:Idea} and \textit{Person:Grounding} were submitted, those entries are highlighted yellow and used to inform the \textit{Approach:Idea} response.
\newline\newline
\noindent

\noindent
\begin{tabular}{c c c}
    \setlength{\fboxrule}{2pt}
    \fbox{
        \parbox[c][2cm][c]{0.12\textwidth}{
            \centering 1. Pictorial spaced repetition learning
        }
    } & 
    \fbox{
        \parbox[c][2cm][c]{0.12\textwidth}{
            \centering 2. Visual storytelling for language acquisition
        }
    } \setlength{\fboxrule}{0.5pt} & 
    \fbox{
        \parbox[c][2cm][c]{0.12\textwidth}{
            \centering 3. Cognitive load theory for efficient memorization
        }
    }
\end{tabular}\\\\
%The user can generate as many ideas as they want, iterate on them and re-brainstorm, or directly update the input box. 
\revision{Franky is interested in exploring both ideas (1) and (2). 
%He is familiar with spaced repetition as a learning theory, but did not consider using storytelling as a method for language acquisition.
He selects (2) first and moves on the ground this idea. 
In the grounding entry (\textit{Approach:Grounding}), DynEx details an approach that connects the Chinese phrases together to generate a narrative and correlates each phrase to an image within the story.
%Franky appreciates these details but is unsure if contextualizing the images and Chinese phrases in a story is most suitable for his learning style. 
%He still wants to explore idea (1), so he saves this grounding in the system by clicking the "Save Version" button (Fig \ref{fig:matrix} - G). 
}
\revision{Franky appreciates these details -- he hadn't considered visual storytelling as an approach before. He then selects (1) \textit{"Pictorial spaced repetition learning"} and grounds the idea. The grounding provides concrete details to the spaced repetition algorithm, such as pairing each image to a word and scheduling review times according to the user's progress. 
After seeing the groundings for both ideas (1) and (2) (the full outputs can be seen in Appendix \ref{app:usage_matrix}), Franky decides to implement (1) \textit{"Pictorial spaced repetition learning"}, since he believes he is better suited to learning the Chinese phrases through memorization, rather than storytelling. He moves on to the Interaction dimension. } 

% The grounding further details when a phrase will be pushed back into the spaced repetition algorithm, and how the application will schedule the algorithm.  

\subsubsection{Interaction}
\revision{Franky generates ideas in the Interaction dimension (\textit{Interaction:Idea}).}
%The entries for the Person and Approach dimensions turn yellow, since they are used as context to generate Interaction ideas. }
%In the Design Matrix entry for , the system brainstorms the high-level interaction paradigm of the application, such as a swipe interface or chatbot. 
\revision{Three ideas appear on in the matrix entry: quiz, flashcard, and game interface, all interaction paradigms that effectively showcase a spaced repetition algorithm. Franky selects the quiz idea -- he doesn't want a game interface, thinks flashcards aren't engaging enough, but believes that quizzes are a quick and effective approach to test language acquisition. He moves on to ground the idea in the final entry of the matrix (\textit{Interaction:Grounding}).}
% \begin{itemize}
%     \item \textit{"Simple guess-and-review quiz interface"}
%     \item \textit{"Visual dictionary flashcard interface"}
%     \item \textit{"Image-based language learning game interface"}
% \end{itemize}
%There exists obvious paradigms for a given approach. In this case,"Visual flashcard interface" and "Game interface is the most obvious for spaced repetition".
%However, it is the more non-obvious ones that are interesting and what the user is intrigued by. 
%The user selects (1) \textit{"Simple guess-and-review quiz interface"} as the idea and moves on to \textit{Interaction:Grounding}. 
%and DynEx system specifies the features and details of the UI components. }
%It determines what information will be displayed, the nature of user interactions, and how these elements will contribute to the overall user experience. 
\revision{The grounding specifies how the quiz interface will be formatted. The quiz will present the Chinese phrase along with its corresponding image and prompt the user to guess its English meaning. The prototype will allow users to view their progress and self-study before the quiz. The full output for the Interaction column can be seen in Appendix \ref{app:usage_matrix}.}
% The grounding is provided:
% \begin{itemize}
%     \item \textit{"The quiz interface should present the Chinese character or word along with its corresponding image}
%     \item \textit{The user then attempts to guess its meaning. If they respond accurately, the item is pushed back into the review cycle based on the SRS algorithm}
%     \item \textit{If the guess is incorrect, the correct meaning is displayed, and the item is scheduled for another review sooner}
%     \item \textit{Users should have a clear view of their progress and a way to navigate to previously learned words for self-study"}
% \end{itemize}
\revision{Franky thinks that the grounding covers all the essential features he needs in a learning application: a way to study, test, and track progress. He names the prototype (Fig \ref{fig:matrix} - H) and submits the matrix.} Now that the Design Matrix has been traversed, he can begin implementation in code. 

\subsection{Code-Generation through Modular Stepwise Implementation}
\begin{figure*}[h]
    \centering
    \includegraphics[width=\textwidth]{figures/system_implementation.png}
    \caption{\textbf{DynEx's Implementation User Interface:} Users can navigate between existing prototypes (A) and create new prototypes (B). Users are suggested and can select from a list of potential project requirements (C). Users are able to edit the project specification (D) and view/modify placeholder data if it is required (E). DynEx breaks down the project specification into implementation steps (F) which can be edited, added, and removed by users (G). Users generate code (H) step-by-step. Users can iterate via natural language (I), toggle between versions (J), and view generated code (K) and interact with UIs (L).}
    
    \Description{Screenshot of DynEx's user interface during the implementation phase, annotated with captions on each side of the interface that call out important features: (A) Navigation bar between existing prototypes, (B) Button to create additional prototypes, (C) Project requirements, (D) Project Specification (spec), (E) Placeholder Data, if required, (F) Implementation Steps, (G) Add, Remove, and Update Step Buttons, (H) Button to generate code, (I) input bar for users to iterate, (J) menu for users to toggle between versions, (K) Generated Code and (L) Window to interact with the UI.}
    
    \label{fig:system_implementation}
\end{figure*}
\subsubsection{Project Scoping}
%-- DynEx will self-invoke DALL-E 2 to generate images for the prototype. 
\revision{To scope the project out before implementation, Franky clicks the "Identify Project Requirements" button (Fig \ref{fig:system_implementation} - C). After a moment, the checkboxes for "Dynamically generated AI-images" and "Pre-generated data" are selected. This makes sense to him -- the pictorial approach he specified in the Design Matrix requires images, and data needs to be generated to populate the Chinese phrases he will learn. Franky has the option to check or uncheck requirements if he wants, but he is satisfied with what DynEx has recommended.}
\revision{Franky then clicks "Generate New Spec" (Fig \ref{fig:system_implementation} - D}), and a spec appears, detailing that the data must contain Chinese words, phrases, and IDs, and specifying that images must be returned for each quiz question to assist with visual learning. It describes 3 sections: Flashcards for self learning, Quiz for knowledge testing, and Progress for the user to track their learning journey (Appendix \ref{app:usage_spec}).
Finally, Franky clicks "Generate Data"\revision{(Fig \ref{fig:system_implementation} - E), and Chinese phrases to practice with appear on the screen (Appendix \ref{app:usage_data}). Franky can regenerate or directly edit the outputs for the spec and the data, but he is satisfied with the overall project scoping. He moves on to generate code for the prototype. }
%, making this step essential for effective app development. 

%Before implementation, the system identifies prerequisite project requirements, generates a (spec), \revision{and generates synthetic placeholder data if needed} (see Fig ~\ref{fig:system_implementation}- C, D, E). The project requirements detail whether or not the prototype requires GPT, images, placeholder data, or supported external libraries. The system brainstorms these requirements using the Design Matrix as context. Users can manually update and change the project requirements if needed - for example, if the application is a book recommendation app, and DynEx did not recommend the images as a requirement, the user can add that requirement themselves if they so please. Using the project requirements and Design Matrix as context, the system then generates a project specification to translate the abstract ideas across each dimension into a technical implementation.

\subsubsection{Modular Stepwise Implementation}

\begin{figure*}[h]
    \centering
    \includegraphics[width=\textwidth]{figures/system_step_outputs.png}
    \caption{\revision{\textbf{Step Outputs}: UI outputs created at each step in the usage scenario. }}
    \Description{Screenshot of UI outputs created at each step of the usage scenario. }
    \label{fig:system_step_outputs}
\end{figure*}

\revision{To begin implementation, Franky clicks "Generate Plan" on the interface, and a step list breaking the spec into multiple steps pops up on the screen (Fig \ref{fig:system_implementation} - F):}
\begin{enumerate}
    \item \textit{Set up the React application and create the main layout with the three sections: 'Flashcards', 'Quiz', and 'Progress'}
    \item \textit{Implement the 'Flashcards' section}
    \item \textit{Implement the 'Quiz' section, utilizing the spaced repetition algorithm}
    \item \textit{Implement the 'Progress' section}
    \item \textit{Call GPT to generate images for the Chinese characters}
\end{enumerate}
\revision{Franky can regenerate, approve, or modify this step list, but he likes this plan and starts to generate code for each step sequentially.  He clicks step (1) \textit{"Create the HTML structure"} to navigate to that part of the implementation phase and clicks the "Generate Code" button (Fig \ref{fig:system_implementation} - H). After a moment, the prototype's UI is rendered for Franky to interact with. He sees a basic HTML, and in the Flashcards section, there is a grid filled with the pre-generated Chinese phrases (Fig \ref{fig:system_step_outputs} - Step 1). He clicks the "Quiz" and "Progress" tab -- the content for these tabs are empty, but since this step only created the HTML structure, that is expected. He can also view or edit the code by clicking the "Code" tab (Fig \ref{fig:system_implementation} - K). Franky can either accept this code or fix it -- he accepts it and proceeds to the next step.

Franky clicks (2) \textit{"Implement the ‘Flashcards section’"} and generates the code for that step. In the Flashcards section, he can now click left and right to swipe through the cards and study the Chinese phrases (Fig \ref{fig:system_step_outputs} - Step 2). He approves step 2, and moves on to step 3.}

After generating code for (3) \textit{"Implement the ‘Quiz’ section,"} Franky discovers that there’s an error with the “submit” button in the Quiz section. The Quiz section shows the Chinese phrase and prompts the user to type in the corresponding English meaning as expected, but after the user submits their answer,
the prototype displays both the next quiz question and the next quiz question’s answer. 
Franky can debug this by iteratively prompting the system in the right direction to regenerate code (Fig ~\ref{fig:system_implementation} - I), redoing the step (Fig ~\ref{fig:system_implementation} - H), or clicking the "Code" tab (Fig \ref{fig:system_implementation} - K) to debug the generated code by hand. 
\revision{Franky makes use of the “Iterate” box and types in the input: “After I click submit, it moves on to the next question but also shows the answer of the next question.”
After re-generating the code, the debugged UI and code is shown on the interface. Franky can iterate as many times as he wants and switch between versions until he is happy with the final step output (Fig ~\ref{fig:system_implementation} - J). Franky tests the regenerated code -- the Quiz section is working as expected, so he approves step 3 and moves on to step 4.

Franky generates the code for (4) \textit{"Implement the ‘Progress’ section"}. In the prototype's UI, he sees a metric for accuracy rate in the Progress section. When he tests the interface, he sees his accuracy rate go up based on how he scores on the Quiz section. Pleased that he can now track his progress, he moves on to step 5.

After generating the code for (5) \textit{"Call GPT to generate images."}, Franky sees that the Quiz section now has images alongside each Chinese phrase. Now, he has visuals to assist him in language learning. The prototype for the original implementation plan is complete.}

\revision{Franky tests the prototype and decides that he would like another feature: in the Progress section, Franky would like a visual indicator to show which words he has mastered. 
He clicks the "Add Step Beneath" button (Fig ~\ref{fig:system_implementation} - G), and types in "Allow users to see Mastered words" in the input box. He then generates the code for this new step and tests the prototype -- all the phrases he has answered correctly in the Quiz section now show up in the Mastered Words section under the Progress tab.

After mastering the initial pre-generated phrases in the prototype, Franky wants a way to expand his initial dataset. He adds a step, types "suggest new phrases for me to learn," and generates the code for that step. Now, after Franky masters his initial set of words, the prototype will use tool-calling to dynamically suggest new phrases with GPT, ensuring that he always will have novel and relevant phrases to learn whenever he wants.
Satisfied that he can continuously expand his language skills, Franky decides his prototype is complete. 
%As a visual learner, he is excited that each phrase is paired to an image and that he can learn phrases via spaced repetition. He also finds the inclusion of a flashcard study section and a progress bar as thoughtful additions to the quiz interface, allowing him to monitor his progress and effectively retain language knowledge. 

}
%Dynex enables users to add more features as they test prototypes. 
%After using the app and testing it out, the user 

% \subsubsection{Creating Multiple Prototypes}
% \revision{Franky wants to explore another prototype, but this time, he wants to try the other Approach that he was curious about when first traversing the Design Matrix, (2) \textit{"Visual storytelling for language acquisition"}, while keeping the dimensions for Person and Interaction the same. He clicks the "+" button at the top right of the interface (Fig ~\ref{fig:system_implementation} - B), which brings him back to the Design Matrix. He can then create a new prototype with this Approach. The different prototypes are displayed in the sidebar of DynEx (Fig ~\ref{fig:system_implementation} - A), allowing Franky to easily revisit any prototype, pick up where he left off, and continue iterating.} 

\section{System Implementation}

The system was implemented in Python, Typescript, and Flask. The Design Matrix uses GPT-4 --- due to its vast understanding across various domains, it is well suited for brainstorming and refining ideas.
DynEx use Claude 3.5 Sonnet for code generation --- Claude is known to be better at writing UI code, and it enables us to use Claude Artifact as a baseline for comparison in our evaluation. Claude is also highly responsive to explicit instructions, which is crucial for code-generation. \revision{The repository is open-sourced on Github.\footnote{https://github.com/jennygzma/ui-design-prototype}}

\subsection{Design Matrix}
\revision{When the user submits a problem, they use the system to brainstorm each entry in the Design Matrix. For each of the six entry types, we provide few-shot examples to guide the type of response desired. 
Three different applications are used as few-shot examples: OKCupid, Tinder, and Coffee Meets Bagel. These applications broadly span the range of the design space for dating apps; while they are all meant for single people, they have different target users, core approaches, and interaction paradigms.
%OKCupid is for users who want to find partners with specific criteria like race and religion (Person), resulting in a searchable database (Approach) and faceted browsing interface where users can filter by criteria (Interaction). Tinder caters to users who value physical attraction (Person) -- it has a card-swipe interface with a large profile photo and brief tagline on each card (Interaction) to lower the cognitive load and make decisions faster (Approach). Coffee Meets Bagel is for users who are looking for serious relationships (Person). Users are presented with a feed of only 5 options (Interaction), which also lowers the cognitive load and forces them to make intentional judgements (Approach). 
Each application example is broken down across dimensions and levels of specificity to establish the desired level of detail for each entry. The full few-shot examples can be seen in Appendix \ref{app:example_matrix}. 

Additionally, the system takes in all previously-submitted entries of the matrix, since they are already-known aspects of the project, to help inform the response of the current entry. The previous entries
used as context are highlighted in yellow on the UI for the user to
visualize. If the user chooses to resubmit a dimension
on the Idea level, the Grounding for that dimension will not be
factored as input context. Ultimately, this ensures that the current design is always factored in when suggestions for each entry, and that the dimensions are synthesized to form a coherent prototype design.
}

\subsection{Code Generation}
% The system first identifies the project requirements, creates the project specification (spec), and generates synthetic placeholder data if required (see Fig ~\ref{fig:system_implementation}- C, D E). Once the initial setup is complete, the user and system collaborate in creating the prototype through step-by-step implementation, where the system breaks down the implementation plan into steps (see Fig ~\ref{fig:system_implementation}- F).

\subsubsection{\revision{Project Scoping}}
% Before implementation, the system identifies prerequisite project requirements, determining whether the prototype requires GPT, images, external libraries, or synthetic placeholder data. It also creates a project specification. 
\revision{The system identifies the technical requirements using the Design Matrix as context, where the contents of the Design Matrix are converted into a natural language prompt detailing what each entry in the Matrix is. The technical requirements and Design Matrix are then used as context to generate the spec. We use few-shot examples to guide the generation, which can be found in Appendix \ref{app:example_spec}. We use the spec, along with few-shot examples, to generate the data, which can be seen in Appendix \ref{app:example_data}. }
% We leverage few-shot examples with LLMs to guide the generation of relevant and appropriate placeholder data. 

\paragraph{External Libraries}
DynEx further enhances prototype functionality with pre-approved libraries. \revision{DynEx's code outputs all use Material UI (MUI), a comprehensive React component library offering foundational elements like buttons, dropdowns, input forms, and more to create visually appealing applications}. Additionally, DynEx's project requirements allow the selection of Chart.js, an open source Javascript library used for creating interactive charts on websites, and GoJs, as Javascript library for creating interactive diagrams such as flow charts, mind maps, and process diagrams. 
%\sout{A common issue with LLM code-generation tools is their difficulty in handling external libraries and packages. From experimentation, OpenHands can get stuck in a loop when installing unusable packages, while tools like Claude Artifact fail when trying to import non-existent component libraries. These limitations pose significant challenges for generating reliable code with LLMs. To prevent this, DynEx only uses the MUI React Library. However, this restriction can limit the complexity of generated applications, so we allow the use of Chart.js and GoJS for creating charts and diagrams.}
When the system recommends these libraries as requirements, we use few-shot code examples to ensure smooth integration.
%This approach ensures safe and reliable code generation, preventing avoidable bugs. 
The list of pre-approved libraries can be expanded this way to ensure safe and reliable code generation. 

\subsubsection{\revision{Modular Stepwise Implementation}}

%To implement the prototype, DynEx employs a \revision{modularized, stepwise} approach to development that allows for human direction at every step, which can be seen in Figure ~\ref{fig:system_implementation_diagram}. 
%The system breaks down the spec into steps, where each step is the next-smallest testable iteration of the previous step.
%The user can regenerate, approve, or modify this step list. 
\revision{DynEx employs few-shot examples to generate the plan, which can be found in Appendix \ref{app:example_plan}. The few-shot examples direct the system to generate a plan that mirrors how a developer would naturally approach the task -- where each step is modular and progressively adds features to the prototype, building on what has already been implemented. }
%\revision{Additionally, our stepwise implementation approach preserves the code at each stage, enabling modularization across steps, and maintaining a clear version history. This ensures that the most advanced working prototype is captured while providing the flexibility to iterate and rollback between steps. 
%}
\revision{
The plan has approximately 3-6 steps, depending on the complexity of the application. At each step, DynEx calls Claude's API to generate code.
To maintain continuity, the prompt explicitly instructs Claude to retain code from previous steps. DynEx also employs a function to clean up any natural language in the code response, ensuring the output is executable. The code rules used in the prompt can be found in Appendix \ref{app:example_code_basic}.
%We specify code rules in the prompt depending on whether or not GPT is invoked, external libraries are used, or synthetic data is generated. 
%At each step, the user can regenerate the code, directly modify the code, or iteratively prompt the generated code in the right direction. Before moving on to the next step, the user must approve the previous step’s code.
}

\paragraph{\revision{Debugging and Iteration within Steps.}}
\revision{
LLM code-generation is not perfect, and there are many ways that DynEx supports error recovery. 
\revision{The user has the option to regenerate the code within each step, manually edit the generated code and preview its output, or rollback to a previous step with working code.
DynEx also offers an “Iterate” box to debug the program by explaining the bug through the prompt. }\revision{The system will reissue a query to Claude, prompting it to fix the buggy snippet of code while keeping the remaining code intact. The prompt can be seen in Appendix \ref{app:example_debug}.}
%All the iterated versions of the debugged code and UI are shown in the system.
%Users can iterate as many times as they want and switch between versions within the step until they are happy with the final step output. 
}

\subsubsection{Self-invoked Multi-Modal LLMs} 
Dynex self-invokes GPT-4 and DALL-E 2 to enable dynamic and flexible content creation based on the project requirements. 
\revision{If GPT or Images is selected as a requirement,
%(see Fig ~\ref{fig:system_implementation}- C), 
we use few-shot code examples to guide the self-invocation. In the few-shot prompt, we provide multiple Javascript code snippets that use a try-catch block to make asynchronous API calls to GPT-4 and DALL-E 2. Each example includes appropriate parameters, such as the OpenAI API endpoint URL, headers for authorization (including the API key), and necessary model identifiers. These snippets guide the generation of code that properly demonstrate how to structure safe OpenAI API calls. Examples can be found in Appendix \ref{app:example_gpt} and \ref{app:example_images}}.

\section{Evaluation}

\begin{figure*}[h]
    \centering
    \includegraphics[width=\textwidth]{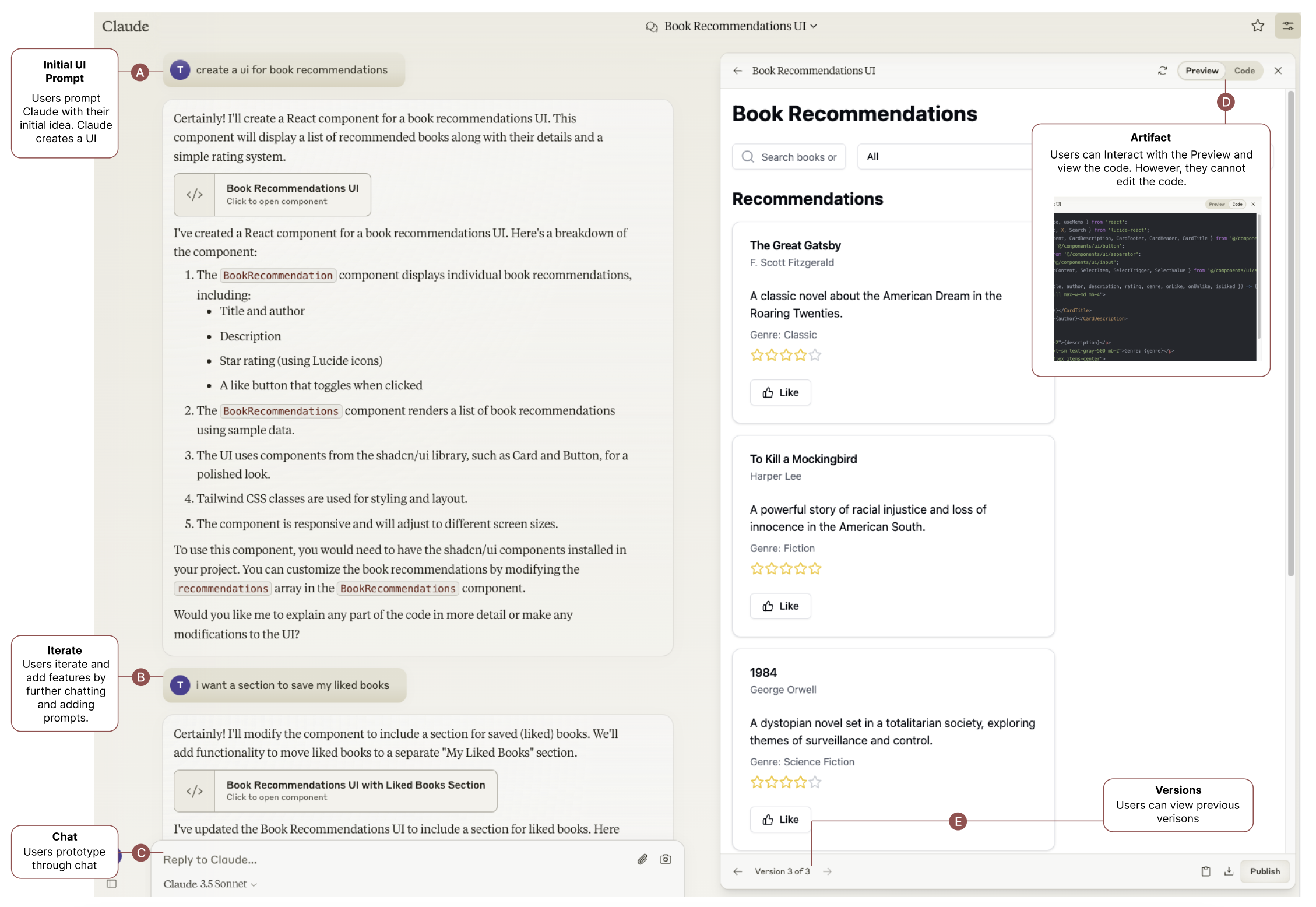}
    \caption{\textbf{Claude Artifact Interface}: Claude Artifact has a chat interface, where users can prompt the chat and have code returned instantly. Users can view the code and the UI on the artifact to the right; users cannot edit the code. Users can iteratively append features to the initial prototype through chat.}
    \Description{
    A screenshot of Claude's user interface during the implementation phase, annotated with captions on each side of the interface that call out important features. Highlighted Features include: (A) user's initial prompt; (B) follow-up prompts iterating on created UIs; (C) chat to send additional prompts; (D) the Artifact, which enables the code preview; (E) version toggle. Users can prompt the chat and have code returned instantly. Users can view the code and the UI on the artifact to the right; users cannot edit the code. Users can iteratively append features to the initial prototype through chat.}
    \label{claudeui}
\end{figure*}

To evaluate DynEx as a system for exploratory programming, we conducted a user study that focused on the following research questions:

\begin{itemize}
  \item \textbf{RQ1: Divergence -} To what extent does DynEx enable divergent exploration within a problem space?
  \item \textbf{RQ2: Convergence -} To what extent does DynEx allow users to better develop their idea?
  \item \textbf{RQ3: Implementation -} To what extent does DynEx enable the code to realize a complex idea? 
  \item \textbf{RQ4: Overall -} To what extent does DynEx allow for a better prototyping experience? 
\end{itemize}

\subsection{Methodology}
\subsubsection{Participants}

Our evaluation was conducted through a qualitative study of 10 programmers (6 male, 4 female); there were 6 industry software developers and 4 CS students.
They were recruited through a snowball sampling of local university students and alumni. Information regarding the participants age, role, and years of technical exposure are included in Table ~\ref{participants}.
Participants were paid \$20 per hour for their time, and the study was conducted with each participant for approximately 90 minutes. 

\begin{table*}[h]
\centering
\resizebox{\textwidth}{!}{%
\begin{tabular}{|p{0.08\textwidth}|c|c|p{0.38\textwidth}|c|}
\hline
\centering{\textbf{ID}} & \textbf{Age} &\textbf{Gender} & \centering{\textbf{Role}} & \textbf{Years of Technical Exposure} \\
\hline
\centering{\textbf{1}} & 20 & M & \raggedright Undergraduate Computer Science Student & 3 \\
\hline
\centering{\textbf{2}} & 25 & F & \raggedright Software Engineer at Large Company & 7 \\
\hline
\centering{\textbf{3}} & 24 & F & \raggedright Software Engineer at Large Company & 6 \\
\hline
\centering{\textbf{4}} & 22 & F & \raggedright Software Engineer at Large Company & 5 \\
\hline
\centering{\textbf{5}} & 25 & M & \raggedright Software Engineer at Large Company & 7 \\
\hline
\centering{\textbf{6}} & 25 & M & \raggedright Software Engineer at Large Company & 7 \\
\hline
\centering{\textbf{7}} & 21 & F & \raggedright Master's Computer Science Student & 5 \\
\hline
\centering{\textbf{8}} & 24 & M & \raggedright Doctoral Computer Science Student & 6 \\
\hline
\centering{\textbf{9}} & 24 & M & \raggedright Doctoral Computer Science Student & 6 \\
\hline
\centering{\textbf{10}} & 26 & M & \raggedright Software Engineer at Start Up & 5 \\
\hline
\multicolumn{4}{l}{}
\end{tabular}
}
\newline
\newline
\caption{Demographics of user study participants. Technical Exposure refers to the number of years since the participant first programmed. }
\Description{5 by 11 table; rows are for the ten participants (and one title row). Columns are ID, Age, Gender, Role, and Years of Technical Exposure.}
\label{participants}
\end{table*}

\subsection{Design Exploration through the Design Matrix}
\subsubsection{Task}
\begin{figure*}[h]
    \centering
    \includegraphics[width=\textwidth]{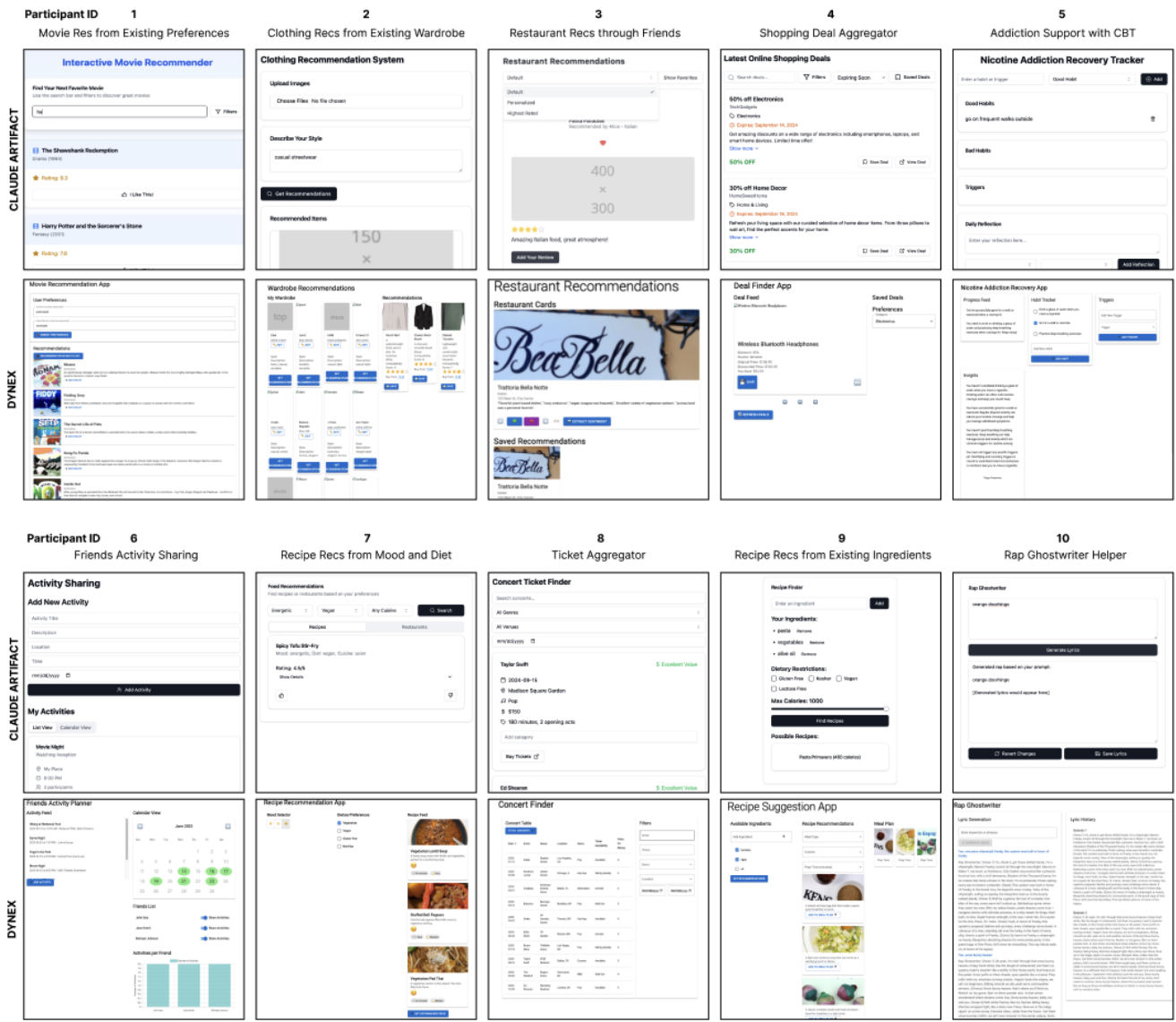}
    \caption{\textbf{Participant Outputs:} UI prototypes created by all 10 participants using both Claude Artifact and DynEx. These UI prototypes spanned a diverse range of use cases from addiction support with CBT to a rap ghostwriter helper. DynEx allowed participants ground prototypes around the real-world constraints of a problem such as the Person, Approach, and Interaction. }
    \Description{
    A screenshot of UIs that the 10 participants created. Each participant created 2 UIs - one with Claude Artifact and one with DynEx. Participant 1 created a movie recommendation system based on existing user preferences. Participant 2 created a clothing recommendation system based on their existing wardrobe. Participant 3 created a restaurant recommendation system through friends. Participant 4 created a shopping deal aggregator. Participant 5 created an addiction support system utilizing concepts of CBT. Participant 6 created a friends activity sharing app. Participant 7 created a recipe recommendation app based on the user's mood and diet. Participant 8 created a ticket aggregator. Participant 9 created a recipe recommendation system based on existing ingredients. Participant 10 created a rap ghostwriter helper.
    }
    \label{fig:participant_outputs}
\end{figure*}

Participants were asked to name a problem that they would like to explore a solution for. The task was to explore concepts and prototypes that would help solve their problem. They were asked to produce a MVP per system. Participants had a broad range of ideas, from recommendation applications for movies, clothing, and restaurants, to a friend-activity sharing app, to a rap ghost-writing assistant (see Fig~\ref{fig:participant_outputs}). They tested two systems: DynEx and Claude Artifact.  

% \begin{table}[h]
% \centering
% \begin{tabular}{|p{0.05\textwidth}|p{0.5\textwidth}|} % Adjusted the second column width to align the table properly
% \hline
% \rowcolor{lightgray}
% \multicolumn{1}{|c|}{\textbf{ID}} & \multicolumn{1}{c|}{\textbf{Idea}} \\
% \hline
% \centering{\textbf{1}} & Movie Recommendations based on Existing Preferences\\
% \hline
% \centering{\textbf{2}} & Clothing Recommendations Based on Existing Wardrobe \\
% \hline
% \centering{\textbf{3}} & Restaurant Recommendations Through Friends\\
% \hline
% \centering{\textbf{4}} & Shopping Deal Aggregator \\
% \hline
% \centering{\textbf{5}} & Addiction Support with CBT\\
% \hline
% \centering{\textbf{6}} & Friends Activity Sharing and Social Management\\
% \hline
% \centering{\textbf{7}} & Recipe Creation based on Mood and Diet \\
% \hline
% \centering{\textbf{8}} &  Ticket Aggregator based on Value-For-Money\\
% \hline
% \centering{\textbf{9}} & Recipe Creation based on Existing Ingredients \\
% \hline
% \centering{\textbf{10}} & Rap Ghost-Writer Assistant\\
% \hline
% \end{tabular}
% \newline
% \newline
% \caption{Participant prototype ideas.}
% \label{topics}
% \end{table}

\subsubsection{Claude Artifact}

We compared DynEx against Claude Artifact, an industry standard tool, as the baseline. Claude Artifact is a chatbot, where every prompt elicits a UI and non-editable code. The chat screen is on the left, and the Artifact to interact with the UI is on the right.
The user can prompt the chatbot to brainstorm more ideas or specify more features -- an example can be seen in Figure ~\ref{claudeui}. 

We considered other tools as the baseline, including OpenHands and GPTPilot. We found OpenHands to be less reliable than Claude Artifact in producing functional prototypes. In our exploration with GPTPilot, we found its user interactions limited to technical specifications. It did not allow for sufficient design space exploration, which is largely emphasized in DynEx. 

%because its ability to use UI component libraries. 

\revision{
We found Claude Artifact to be the best available baseline for our evaluation.}
Because Claude Artifact can generate functioning code from short prompts, it is a very powerful tool for prototyping. Claude Artifact is also widely accessible and creates visually-appealing UIs.
Although it is limited to 4096 output tokens (approximately 450 lines of code), it still produces rich prototypes and stores long chat histories that track how the user further developed their application from the initial idea.
\revision{
While it lacks an explicitly structured framework for design exploration, Claude Artifact has all the capabilities for people to explore the dimensions of a design space before generating their code.
Its organic structure empowers users to iteratively refine their designs, and identify and address gaps in their UI prototypes as they emerge. 
This reactive exploration mirrors the natural way people interact with LLMs and offers an intuitive way to discover new possibilities within a design space. 
}
%Since the user has the ability to engage with design space exploration through the flexibility of the chatbot interface, we decided Claude Artifact was the most appropriate baseline for our study.

\subsubsection{Procedure}
An experimenter first explained the concepts of DynEx through a slide deck that introduced design exploration and exploratory programming, focusing on user-centric design principles and LLM code-generation. The experimenter then gave a demonstration of both systems. 
For Claude Artifact, the experimenter demonstrated how to prompt it to create a UI and add features through the chatbot. 
For DynEx, the experimenter demonstrated how to use the Design Matrix and iteratively implement, debug, and add features to their prototype. 
After the demonstrations for each system, participants were given 30 minutes to create a prototype using DynEx and \revision{30 minutes to create a prototype using} Claude Artifact. Half of the participants started prototyping with Claude Artifact, and the other half started with DynEx. \revision{Participants were not allowed to modify the spec, because the spec is a natural translation of the Design Matrix into a technical implementation description.} After the conclusion of the two tasks, an experimenter conducted a semi-structured interview to understand participant experiences.

\subsubsection{Ratings and Interviews}
After using both systems, participants were asked to rate their experience across a 1 (bad) to 7 (good) scale for 10 questions. The first 4 questions were directed towards the exploratory programming experience. The last 6 questions measured the participant workload while using each system, using the NASA Task Load Index (NASA-TLX) ~\cite{hart1988development}. NASA-TLX can assess the overall success of a system because it provides a comprehensive measure of cognitive workload, which directly influences user performance and satisfaction. Users were also interviewed afterwards about their experience with both systems and their thoughts on the prototypes generated.

\subsection{Results}

\subsubsection{RQ1: Divergence - To what extent does DynEx enable divergent exploration within a problem space?}

Our evaluation showed that participants found it much easier to explore different design ideas when using DynEx as compared to a baseline Claude Artifact. Across 10 participants, DynEx was rated significantly higher than Claude Artifact in enabling design exploration (average scores of 6.1/7 and 4.0/7, respectively; see Table~\ref{EP} or Figure~\ref{EP-Graph}). 8/10 participants rated DynEx at least a 6/7 in regards to this metric. A paired t-test shows that difference in ratings of the two systems is statistically significant at the p = 0.05 level (see Table~\ref{EP}).

Participants reported that DynEx’s system-led brainstorming was more effective in exploring different possible solutions. All participants noted that the system articulated required components of the application they recognized, or interesting possibilities they had not considered. For example, P8 stated, \textit{"I didn't start with that many ideas to begin with on my own. [Without DynEx] I really wouldn't have thought of these steps that fast. It was predicting more things than I had imagined."} All participants similarly expressed positive sentiment about being presented with ideas while using DynEx, feeling that using Claude Artifact required significantly more mental exertion to come up with such ideas themselves. \textbf{From these results, we conclude that DyNex enables users to explore a broader range of solutions to their problem.}

\subsubsection{RQ2: Convergence - To what extent does DynEx allow users to better develop their idea?}

We found that participants were able to better develop their initial idea while using DynEx as compared to a Claude Artifact baseline. Across 10 participants, DynEx was given an average score of 5.9/7 for convergent thinking, compared to an average score of only 3.4/7 for Claude Artifact (see see Table~\ref{EP} or Figure~\ref{EP-Graph}). 7/10 participants rated DynEx at least a 6/7 for convergent thinking. A paired t-test shows that the difference in ratings of the two systems is statistically significant at the p = 0.05 level (see Table~\ref{EP}).

Participants indicated that they were able to sufficiently build off of their own ideas using DynEx. P9 stated that DynEx allowed them to \textit{"incorporate [their] initial idea and helped [them] refine it."} Participants also reported that using DynEx’s Design Matrix allowed them to better specify the problem that they were trying to solve, specifically noting that the Person dimension was helpful to inform the remaining entries of the matrix. Participants also felt that they were better able to explore the solution space with DynEx as compared to Claude Artifact. 
%Most expressed that DynEx provided them with options on how to solve the problem, whereas Claude Artifact was limited to implementing features the participant themselves had to come up with. 
Overall, DynEx was very helpful during the design process; P8 stated that they \textit{"didn't think many [questions were] left unanswered."} \textbf{From these results, we conclude that DynEx is more successful at developing users ideas.}

\subsubsection{RQ3: Implementation - To what extent does DynEx enable the code to realize a complex idea? }

\begin{figure}[h]
    \centering
    \includegraphics[width=0.5\textwidth]{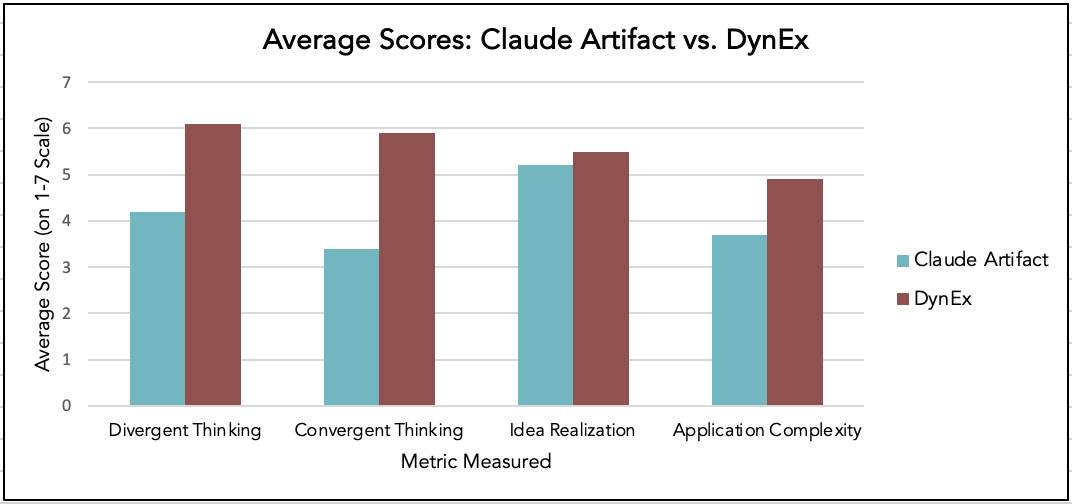}
    \caption{Average Scores for Claude Artifact and DynEx for exploratory programming metrics. The difference in averages is statistically significant for Divergent Thinking, Convergent Thinking, and Application Complexity.}
    \Description{Bar Graphs comparing average participant scores for each of the four exploratory programming questions. DyNex outperforms Claude Artifact across all four questions.}
    \label{EP-Graph}
\end{figure}

Our evaluation demonstrated that participants were able to create more feature-rich and complex solutions using DynEx as compared to Claude Artifact. The 10 participants scored both systems similarly in terms of code-generation realizing their ideas 
%realizing an idea via an implementation (5.5/7 and 5.2/7 for DynEx and Claude Artifact, respectively; 
(see Table~\ref{EP} or Figure~\ref{EP-Graph}), but there was a major difference in participant scores regarding the complexity, which we define as the feature-richness of applications, that were created. Participants scored DynEx (average of 4.9/7) higher than Claude Artifact (average of 3.7/7) in terms of the complexity of the application produced (see Table~\ref{EP} or Figure~\ref{EP-Graph}), with several describing their DynEx prototypes as being \textit{"more feature-rich"} than their Claude Artifact counterparts. 8/10 participants scored DynEx prototypes as being at least a 5/7 in terms of complexity, whereas only 3/10 participants scored their Claude Artifact prototypes above 5/7. A paired t-test shows that the difference in ratings of the two systems in terms of complexity is statistically significant at the p = 0.05 level (see Table~\ref{EP}). \textbf{From these results, we conclude that DynEx is able to create more complex and feature-rich applications.}

\begin{table*}[h]
\centering
\begin{tabular}{|p{0.5\textwidth}|c|c|c|c|}
\hline
\centering{\textbf{Exploratory Programming}} & \textbf{Claude Artifact} & \textbf{DynEx} & \textbf{p-value} & \textbf{t-statistic} \\
\hline
\textbf{Divergent Thinking:} To what extent were you able to explore different conceptual ideas? using \textit{system X}? & 4.2 & 6.1 & \textbf{0.0066} & -3.52 \\
\hline
\textbf{Convergent Thinking:} To what degree were you able to better develop/specify your idea using \textit{system X}? & 3.4 & 5.9 & \textbf{0.00015} & -6.23 \\
\hline
\textbf{Idea Realization:} To what degree was the code able to realize your idea when using \textit{system X}? & 5.2 & 5.5  & 0.678 & -0.43 \\
\hline
\textbf{Application Complexity:} How complex is the prototype produced by \textit{system X}? & 3.7 & 4.9 & \textbf{0.024} & -2.71 \\
\hline
\multicolumn{3}{l}{}
\end{tabular}
\newline
\newline
\caption{Average Participant Score for Exploratory Programming Metrics Between Claude Artifact and DynEx. For all questions, a score of 7 is best (i.e., for all comparisons, a higher number is better). The results from paired t-tests are also listed; statistically significant p-values (p < 0.05) are in bold.}
\Description{5x5 table displaying average score given by participants for either system for four questions (one title row) measuring the system's ability to support exploratory programming. Each question was rated on a scale of 1 (Bad) to 7 (Good). The four questions are 1. Divergent Thinking: To what extent were you able to explore different conceptual ideas? using system X?. 2. Convergent Thinking: To what degree were you able to
better develop/specify your idea using system X? 3. Idea Realization: To what degree was the code able to realize your idea when using system X? 4. Application Complexity: How complex is the prototype
produced by system X? Columns are: the question, average score for Claude Artifact, average score for DynEx, p-value from paired t-test on difference of system scores, t-statistic from the same test. The differences for the first, second, and fourth questions are statistically significant. The difference for the third question is not.}
\label{EP}
\end{table*}

% \begin{table}[h]
% \centering
% \begin{tabular}{|p{0.3\textwidth}|c|c|}
% \hline
% \rowcolor{lightgray}
% \centering{\textbf{Exploratory Programming}} & \textbf{p-value} & \textbf{t-statistic} \\
% \hline
% Divergent Thinking & \textbf{0.0066} & -3.52 \\
% \hline
% Convergent Thinking & \textbf{0.00015} & -6.23 \\
% \hline
% Idea Realization & 0.678 & -0.43 \\
% \hline
% Application Complexity & \textbf{0.024} & -2.71 \\
% \hline
% \end{tabular}
% \newline
% \newline
% \caption{Statistical Significance of the difference between scores for Claude Artifact vs. DynEx. Statistically significant p-values (i.e., p < 0.05) in \textbf{bold}. }
% \label{EP-Stats}
% \end{table}

\subsubsection{RQ4: Overall - To what extent does DynEx allow for a better prototyping experience?}
Overall, we found that DynEx was more successful for exploratory programming compared to Clauade Artifact. 
%Our evaluation of participants NASA-TLX metric scores for both systems suggested that DynEx offers a better user experience while prototyping as compared to Claude Artifact. 
In terms of success in accomplishing the task at hand (NASA-TLX performance), participants on average scored DynEx 5.9/7 compared to only 4.2/7 for Claude Artifact (see Table~\ref{TLX}). 8/10 participants rated DynEx a 6/7 or higher in terms of NASA-TLX performance. This difference in rating was statistically significant at the p = 0.05 level (see Table ~\ref{TLX}).

The difference in ratings between DynEx and Claude Artifact for the remaining NASA-TLX metrics (mental demand, physical demand, temporal demand, effort, and frustration) were statistically insignificant (see Table ~\ref{TLX}). However, there were still interesting findings for the mental effort and demand metrics. These NASA-TLX scores had a large range of responses; many participants found that DynEx gave them freedom to think conceptually, but also required them to read and wait longer compared to Claude Artifact. However, users expressed that Claude Artifact required more mental exertion to brainstorm ideas and guide the prototype's development direction. Hence, depending upon user's preferences (i.e., do they prefer reading LLM-generations or writing prompts themselves), scores for mental exertion may vary. This same discrepancy can explain the lack of statistical significance in the difference in ratings for the NASA-TLX effort category. 
%The difference in NASA-TLX ratings for temporal demand were likely insignificant because participants were given 30 minutes to create one prototype with the system, which was likely more-than-enough time. We did not expect participants to be more frustrated or feel more physical exertion with either system; hence, the lack of difference between ratings for these metrics is unsurprising. 
\textbf{From these results, we conclude that DynEx more successfully allowed users to create a final application as compared to the Claude Artifact baseline.}

% \begin{table}[h]
% \centering
% \begin{tabular}{|p{0.5\textwidth}|c|c|}
% \hline
% \rowcolor{lightgray}
% \centering{\textbf{Mental Workload/NASA-TLX}} & \textbf{Claude Artifact} & \textbf{DynEx} \\
% \hline
% \textbf{Mental Demand:} How mentally demanding was the task when using \textit{system X}? & 3.2 & 4 \\
% \hline
% \textbf{Physical Demand:} How physically demanding was the task when using \textit{system X}? & 5.7 & 5.7 \\
% \hline
% \textbf{Temporal Demand:} How hurried or rushed was the pace of the task when using \textit{system X}? & 4.7 & 5.4 \\
% \hline
% \textbf{Performance:} How successful were you in accomplishing what you were asked to do when using \textit{system X}? & 4.2 & 5.9 \\
% \hline
% \textbf{Effort:} How hard did you have to work to accomplish your level of performance when using \textit{system X}? & 4 & 4.8 \\
% \hline
% \textbf{Frustration:} How insecure, discouraged, irritated, stressed, and annoyed were you when using \textit{system X}? & 4.6 & 4.8 \\
% \hline
% \end{tabular}
% \newline
% \newline
% \caption{Average Participant Score for Mental Workload/NASA-TLX Metrics Between Claude Artifact and DynEx. For all questions, a score of 7 is best (i.e., for all comparisons, a higher number is better). This includes questions such as mental/physical demand - a 7 indicates that there was low exertion. }
% \label{TLX-Averages}
% \end{table}

\begin{table*}[h]
\centering
\begin{tabular}{|p{0.5\textwidth}|c|c|c|c|}
\hline
\centering{\textbf{Mental Workload/NASA-TLX}} & \textbf{Claude Artifact} & \textbf{DynEx} & \textbf{p-value} & \textbf{t-statistic} \\
\hline
\textbf{Mental Demand:} How mentally demanding was the task when using \textit{system X}? & 3.2 & 4 & 0.280 & 1.15 \\
\hline
\textbf{Physical Demand:} How physically demanding was the task when using \textit{system X}? & 5.7 & 5.7 & 1.0 & 0.0 \\
\hline
\textbf{Temporal Demand:} How hurried or rushed was the pace of the task when using \textit{system X}? & 4.7 & 5.4 & 0.242 & 1.25 \\
\hline
\textbf{Performance:} How successful were you in accomplishing what you were asked to do when using \textit{system X}? & 4.2 & 5.9 & \textbf{0.003} & -4.02 \\
\hline
\textbf{Effort:} How hard did you have to work to accomplish your level of performance when using \textit{system X}? & 4 & 4.8 & 0.196 & 1.39 \\
\hline
\textbf{Frustration:} How insecure, discouraged, irritated, stressed, and annoyed were you when using \textit{system X}? & 4.6 & 4.8 & 0.785 & 0.28 \\
\hline
\end{tabular}
\newline
\newline
\caption{Average Participant Score for Mental Workload/NASA-TLX Metrics for Claude Artifact and DynEx. For all questions, a score of 7 is best (i.e., for all comparisons, a higher number is better). This includes questions such as mental/physical demand - a 7 indicates that there was low exertion. The results from paired t-tests are also listed; statistically significant p-values (p < 0.05) are in bold.}
\Description{5x7 table displaying average score given by participants for either system for six questions (one title row) measuring the workload required to use either system based on NASA-TLX Metrics. Columns are: the question, average score for Claude Artifact, average score for DynEx, p-value from paired t-test on difference of system scores, t-statistic from the same test. Only the difference for the fourth question (performance) was statistically significant.}
\label{TLX}
\end{table*}

% \begin{table}[h]
% \centering
% \begin{tabular}{|p{0.2\textwidth}|c|c|}
% \hline
% \rowcolor{lightgray}
% \centering{\textbf{NASA-TLX}} & \textbf{p-value} &\textbf{t-statistic} \\
% \hline
% Mental Demand & 0.280 & 1.15 \\
% \hline
% Physical Demand & 1.0 & 0.0 \\
% \hline
% Temporal Demand & 0.242 & 1.25 \\
% \hline
% Performance & \textbf{0.003} & -4.02 \\
% \hline
% Effort & 0.196 & 1.39  \\
% \hline
% Frustration & 0.785 & 0.28  \\
% \hline
% \end{tabular}
% \newline
% \newline
% \caption{Statistical Significance of the difference between scores for Claude Artifact vs. DynEx. Statistically significant p-values (i.e., p < 0.05) in \textbf{bold}.}
% \label{TLX-Stats}
% \end{table}

\subsection{General Qualitative Findings }
%In general, participants reported a positive experience while using DynEx. Participants expressed enthusiasm towards 1) DynEx's Design Matrix in the articulation of ideas, both new and familiar, 2) DynEx's ability to realize prototypes with intuitive features and details. 

\subsubsection{DynEx's Design Matrix inspired new solutions}
DynEx's Design Matrix enabled participants to consider ideas they had not thought of, broadening their initial concept and allowing them to adequately explore the design space. 
%P10 noted that \textit{"without knowing what I wanted, it did a very good job of guessing and from there I was able to branch off and have my own ideas"}. 
P3 prototyped an app that recommended restaurants based on friends reviews. She noted: \textit{"It was really cool that it gave me ideas --- otherwise I would have gone with what I had imagined ... [initially, which was not as good]"}. She had pictured a typical, feed-like-layout, but through DynEx's exploration, ended up prototyping a card-swipe interaction that she enjoyed more. 

P5 experienced something similar, stating that \textit{"For every [entry in the matrix], it included [ideas] that I wanted, but also considered ideas that I hadn't considered but wanted to build...it did a good job of articulating ideas that I had in my mind, touching on tangential pieces that I might be interested in but didn't know how to bring together."} P5 created a addiction recovery application, and was particularly intrigued by using cognitive behavioral theory (CBT) principles brainstormed in the Approach dimension. He ultimately created an app that allowed him to journal triggers, relapses, and good days, utilizing CBT help identify unhelpful thoughts. 

P8 created a prototype to aggregate concert tickets based on value-for-money --- he had been thinking about this idea for the past two years, but the Design Matrix was able to help him make progress with new insights. He stated,\textit{"even though I kind of knew what I wanted to do, I really wouldn't have thought of these steps so fast. Stuff like: 'do you want a centralized database?'... [for even] simple things like that, I would have to think for a much longer time if I wanted to build it myself... it saved me from spending a lot of time on unimportant things... and gave me conceptual freedom."}. 

On the contrary, Claude Artifact did not facilitate the exploration of the design space, but rather, implemented prototypes exactly as the user prompted. P2 noted that \textit{"with DynEx, there's a lot more time put into the [end-user of the prototype] and [a lot more] feature-oriented thoughts...[with Claude], I didn't know what to [prompt] at times... since there were no recommendations."} Additionally, P4 stated that when using Claude Artifact, \textit{"I was just thinking on my own... it did not come up with any of the ideas, I came up with all the ideas"}. The Design Matrix gave users a better starting foundation; P9 stated that \textit{"[DynEx helped me] incorporate my initial idea and refine it... the brainstorming helped me think through rough points [he] had missed out on"}. Exploring the design space prevented users from design fixation, allowed them to ideate on new solutions, and challenged their existing thinking. 

\subsubsection{DynEx's prototypes are more feature-rich and intuitive}
Users were also pleasantly surprised by DynEx's ability to realize more intuitive and realistic applications. DynEx was able to generate more relevant placeholder data and emulate a true user experience. P2 noted that \textit{"having the faked data really helps with understanding what is going on with the system and interacting with it"}, compared to Claude Artifact's \textit{"very simple"} prototype with limited placeholder data. P3, who created the restaurant recommendation app, noted that she could not emulate a user experience from her prototype with Claude Artifact, stating \textit{"its data could not mimic it... [it didn't even] have reviews"}; in comparison, DynEx's placeholder data \textit{"helped a lot [to envision] the UI properly"}. 

DynEx's ability to consider many small details that worked together cohesively also helped realize more thoughtful and intuitive prototypes. P6 had a vague application idea when initially prototyping -- a friend-activity sharing app where friends could join other friend's plans. He eventually settled on a calendar-like interface, but was still surprised by the comprehensive feature-set the prototype provided, stating, \textit{"it suggested many features to answer the foundational set of questions that were trying to be answered, such as how to match people with friends, how to visualize the friend-sharing, joining friend-sharing experiences... it created a pretty robust social ecosystem surrounding the calendar experience."} He felt as though the final prototype \textit{"incorporated really different, very distinct UI features that weren't very connected to each other [at face value] very well... it solved the problems it [set out to solve]."} He indicated interest in using this initial prototype as a starting point for a real, production application for himself to build off of. 

P8 also noted the intuitive features in his concert aggregator prototype. With DynEx, the UI was formatted like a table, with important information surfaced --- the main user interaction prioritized searching for valuable concert tickets with low cognitive load. In contrast, the UI for Claude Artifact had a feed view, with lots of information about each concert and ticket, which P8 disliked after interacting with the prototype. P8 greatly preferred DynEx's prototype, stating: \textit{"It was intuitive. It was feature-rich. It had all the important features, like sorting by columns and sorting by genre, dates, value-for-money etc...I liked the conciseness of the information... contrary to [Claude Artifact's] prototype which was... not-fit for this use case."}. Ultimately, users were satisfied with the cohesion between features in DynEx-created prototypes.

\subsubsection{Claude's immediate feedback loop is engaging}

%Although DynEx was more successful in enabling participants to explore a design space and realize complex and feature-rich prototypes, it 

Users viewed Claude Artifact's responsiveness favorably; they were able to quickly generate a UI based on limited prompts. P4 liked that Claude Artifact \textit{"showed the prototype instantly"}. P5 noted: \textit{"Claude is a chat, which is effortless... because it didn't force me to explore more, it didn't feel mentally demanding."} Similarly, P9 stated that Claude Artifact's \textit{"translation from prompt to output to previewing code is very clear and fast."} Users disliked Dynex's slower code-generation and the dense-text they were required to read through when traversing through the Design Matrix. P9 said that DynEx required \textit{"a bit more mental effort because you had to think through...navigate... and read through the matrix"}. Many users also commented on the speed of the code generation; P1 stated that DynEx was \textit{"very slow"} compared to Claude Artifact. Almost all participants appreciated the speed of Claude Artifact compared to DynEx, emphasizing the importance of an immediate feedback loop. 

However, participants did also note that this was a drawback in terms of enabling a thorough design space exploration --- because of Claude Artifact's inherently linear workflow, users only appended features to their initial prototype, as opposed to trying out different ideas. P4 noted: \textit{"[Claude Artifact] just did what I told it to do... [the prototype] was what I expected"}. P8 stated: \textit{"I couldn't even brainstorm the fact that it could brainstorm,"} reflecting a broader trend among participant usage patterns with Claude Artifact; participants did not spend much time ideating with Claude Artifact because it was not intuitive on how they could do so.

\subsubsection{DynEx bridges the gap between design and engineering}
Participants appreciated the broader context and conceptualization provided by the Design Matrix.
%, likening its function to the role a product manager between design and engineering teams in industry. 
Participants in our study had significant technical experience but limited design experience, with 6/10 currently working as software engineers in industry. P2 stated: \textit{"When it asks who is the impacted user, [it is like] what our design person always asks."} P8 said: \textit{"I appreciated that [DynEx] walked me through the actual entire design experience, similar to how a product manager thinks about a business question - thinking about who's the user? What are the user stories surrounding the users? And then mapping those user stories to actual features."} They appreciated the contextual understanding about the design they gained through traversing the Design Matrix. 
%\newline\newline
%\textbf{Qualitative findings showed that DynEx allows the user to explore a design space, broaden their initial ideas, and create feature-cohesive prototypes that adequately addressed their use case.}

\section{Discussion}

%DynEx successfully enabled exploratory programming, facilitating the transformation of abstract ideas into concrete prototypes. 
%by guiding users to explore their problem space to create complex and feature-rich applications. 
%We discuss how LLMs can significantly enhance and accelerate the process of exploratory programming. We additionally consider the implications of unstructured and structured approaches towards design exploration. Finally, we discuss the potential of the Design Matrix acting as a liaison between designers and engineers.

%\vivian{cutMany individuals, from students to professors to startup founders, have innovative ideas. However, initial ideas are rarely fully thought out - they require exploration to gain deeper insights. Without exploration and development, even the most innovative ideas can falter before they have a chance to have a lasting impact. }

\subsection{LLM Techniques to Enhance Exploratory Programming} 

One of the biggest challenges when using LLMs for design exploration is guiding users towards adequate problem specifications. DynEx was able to address this by providing users with a framework for structured exploration of a design space based on specific dimensions (Person, Approach, and Interaction). In many instances, 
%users were able to select ideas that aligned with their thinking;
%Users primarily built prototypes for themselves. In the Person dimension, 
% In exploratory programming, identifying the key parameters and inputs that form the basis of an idea is crucial to instantiate exploration. --what does this mean - vivian
DynEx offered suggestions users intuitively understood but could not fully articulate, supporting users with recognition over recall. Exploring dimensions of the design space with DynEx also broadened the scope of the exploration process. 
%enabling the creation of variations through repeated changes to parameters and inputs necessary for exploratory programming. 
Our dimensions provided natural parameters that users could repeatedly change to create variations of prototypes. Ultimately, the true power of exploration through dimensions lies in helping users specify their problems, clarifying their design goals, and broadening their landscape of solutions.

LLM code-generation abilities can transform the exploratory programming process. Exploratory programming systems must consider the ease or difficulty of exploration when prototyping. With LLMs, users can prototype UIs rapidly without compromising the complexity of the application through self-invoking multi-modal LLMs. While prototypes should be simple, they need a minimum level of complexity in order to properly mimic the user experience. The self-invocation of generative AI allowed users to better simulate a real application through images, placeholder data, and dynamic data generation. It is a powerful idea for LLMs to leverage their own advanced capabilities to create more functional applications
%. Users appreciated DynEx's ability to self-invoke multi-modal LLMs 
-- without it, recommendation systems wouldn't be able to recommend anything and applications with a visual component of the experience would have no images to flight-test the experience.

To enable users to explore the design space even more thoroughly, we can incorporate dimensions past the Person, Approach, and Interaction, as well as levels of specificity beyond Idea and Grounding. For example, a new dimension that could be considered is multiple stakeholders -- a marketplace application must have functionalities for both buyers and sellers. Another potential dimension is existing solutions -- if designing an application for a problem for which other solutions already exist, users should be able to identify deficiencies in existing solutions to inform the development of their own prototypes. Adding dimensions leads to more thoughtful application designs that take into account more aspects of the problem. % The concept of the Design Matrix can be expanded vertically across levels of specificity as well. For DynEx, each dimension was explored through Idea and Grounding -- w
Additionally, we could add levels of specificity outside of the Idea and Grounding levels, such as a row for mock-ups that provide a visual of the dimension. Adding levels of specificity to further crystallize concepts can result in the creation of a more-detailed and cohesive application.
% This allows users to further specificity concepts per dimension. 

%For example, dating applications like Bumble, Coffee Meets Bagel, and Tinder each cater to different audiences and needs. These apps facilitate connection through different interaction mechanisms: through swipe, through who is allowed the first move initiates, and through what filters they have access to. Applications are created around the problem of a specific intended user and are contextually grounded in the constraints of their target audience. If a user wanted to create a new dating app, they would likewise need to examine these existing solutions, identify gaps for their use case, and understand how their new approach would stand out. Broader dimensions allow users to consider more practicalities and ensures that the design aligns with all stakeholders. 
%A level of specificity further than Grounding that could be considered could be Visual Mockups. 

%\todo{what metric would we evaluate this by? how can they be built into the application?      Dating app example does not fit, needs an example where there are more stakeholder          we can add rows too not just columns}

% , by powering the  with LLMs to brainstorm and refine ideas could lead to even more innovative solutions.

\subsection{Structured vs Unstructured Design Exploration}

While users felt supported by DynEx’s structured design space exploration, they also appreciated aspects of Claude’s chat-interface --- namely, that it provided a more immediate feedback loop. Claude Artifact's interactions are simple and direct --- users could type an input and almost instantly receive a code output. This responsiveness engaged users, motivating them to iterate more quickly on ideas. Chatbot systems are very common; people keep making them, and users are very familiar with them. They invite open-ended input, allowing users be highly specific or intentionally vague in their prompts. However, just because it is common, does not mean it is the best paradigm.  Comparing it effectiveness to a structured UI, such as DynEx, is also important. 

%it certainly had a magical feel to it when interacting with the system that participants found motivating.  
%Because Claude Artifact was able to produce immediate working code through the chat interface,

We found that users do not really use Claude Artifact to explore a problem space, despite the fact that they could prompt the chat to help them explore. Most participants instead worked through a linear process, appending features to Claude Artifact's initial design. Linear processes are known to be problematic because they do not facilitate an exploration of the broader design space, limiting the complexity and intuitive way that features can work together in the application. In fact, all users were unaware or did not even think of utilizing Claude Artifact for exploration outside of feature-addition purposes. 
%was certainly more helpful in creating comprehensive and feature-rich prototypes, and we argue, essential for true exploratory programming. 

%Yet, exploratory programming also values the ease of exploration--the key question is 
It remains an open question as to exactly how structured design space exploration should be without compromising the implementation process. DynEx's Design Matrix required users to explore the problem space before implementation, but users may not always be ready to engage with the detailed considerations that structured exploration demands.
%and can feel intimidated by it. 
%It's obvious that LLMs can assist with code generation, but their role in structured design exploration is less clear. 
Future exploratory programming tools should consider ways to combine the immediate feedback and flexible prompting of Claude Artifact with the comprehensive design exploration of DynEx. Perhaps a user could begin with with unstructured exploration, and once sufficiently inspired, transition to a more structured approach to tackle more complex solutions. Designing prototyping tools that can strike a balance between ease and structure could encourage more individuals to explore unrealized ideas, foster innovation, and solve real-world problems.

\subsection{\revision{Bridging the gap between design and engineering}}

\revision{DynEx brings together design and implementation through the Design Matrix and code generation. 
Currently, DynEx is built for individuals, but in the working world, designers and engineers usually work in teams.}
%While DynEx brings together design and programming for individuals with successful results, in the working world, designers and engineers work in teams.
Traditionally, designers and engineers are siloed during software development - designers design and engineers implement, with product managers attempting to bridge the gap between the two. Context is bound to be lost when requirements are handed off across teams.
%However, the design landscape is always changing, and context is bound to be lost -- even a product manager may not be up to date on the latest changes. 
Our user study with individuals demonstrated the value in contextualizing implementations in design, and these insights can be scaled to a team environment. 
%Addressing this disconnect between design and engineering is challenging but useful;  
%However, in the working world, designers and engineers work in teams. Designers must continuously update engineers through shared Figma files as designs change.
\revision{
%Incorporating design context into these LLM-powered systems could extend to tools for software development workflow on teams.
%Designers must constantly update engineers as designs evolve, often relying on tools like shared Figma files; 
The Design Matrix can bridge the gap between teams by providing a continuous feedback loop between designers and engineers. 
Design and engineering teams can contribute their unique perspectives and expertise in the matrix: 
%where different teams can contribute their expertise for collaborative input. 
designers can populate the Person through user interviews, backend engineers can define the Approach, and front-end engineers can describe the Interaction. 
This shared structure ensures that all perspectives are synthesized and valued, reducing inefficiencies from conflicting opinions and redundant meetings that often lack clear, actionable outcomes. 
It also offers a clear reference point for organizations to understand the relationship between design and code.
%A Design Matrix completed by various teams can be incorporated into tools that support code synthesis in established codebases, such as Gemini and CoPilot, to provide contextual awareness between application design and implementation at an organizational level. 
Ultimately, the Design Matrix offers a more collaborative development process, bridging silos and providing context across teams.
}
%Existing LLM-powered tools for established codebases, such as CoPilot and Gemini, only support developers in implementation. 

% Existing LLM-powered tools for established codebases, such as CoPilot and Gemini, only support developers in implementation. Hence, it is necessary to extend DynEx's principles for exploratory programming into workflows involving multiple team members.

%designers want to be engineers and engineers want to be designers
%The design matrix enables designers to play the role of programmers, and
%rappers want to be hoopers and hoopers want to be rappers

% it should not be about dynEx, it should be about GENERALLY to what degree should it help to mix design with prototyping? it is a constant challenge that teams face and they do waterfall and alot of redundant work when in reality that there are design options that are never settled.
\section{Limitations and Future Work}
Our user study was limited to ten participants, all of whom had computer science backgrounds and many years of technical experience. This may not be representative of the broader population that could benefit from an exploratory programming system --- namely, anyone with an idea to prototype, regardless of their past technical exposure. Future studies should expand the demographics of participants to best inform the development of a system for users of varying technical backgrounds. Our study was also limited in the amount of time participants had to prototype. They had 30 minutes with each system --- with additional time, participants may have created a variety of prototypes and compared them with one another to emulate a richer prototyping experience. Future studies should 
% expand the time horizon both for creating one prototype or be extended to longitudal studies, 
examine how users interact with the system when given more time.

Our user study also only used Claude Artifact as a baseline. Although we conducted a brief exploration which suggested Claude Artifact as being the most suitable as a baseline compared to other existing LLM-based tools, future work could involve comparisons between more systems to identify specific components or design choices that are effective in supporting exploratory programming. Furthermore, while users were given an introduction to Claude Artifact, they were still quite inexperienced; experienced users might use Claude Artifact as a tool for prototyping differently than novices. 

%Future work can enable this capability, to observe the ways in which user ideas change after seeing created UIs.

Our system uses Claude 3.5 Sonnet for coding, and while very powerful, it introduced certain limitations. Most notably, Claude has a token cap of 4096 tokens, where API calls can only return approximately 450 lines of code. 450 lines of code is enough to prototype most applications; however, users were limited in complexity when they attempted to add features. Additionally, users expressed dissatisfaction with DynEx’s speed, noting that Claude Artifact offered a faster feedback loop. API responses from LLMs often return undefined code, requiring extra steps for cleaning and debugging the code in our system, which slowed down the prototyping process. As LLMs improve, these "clean-up" steps will also become unnecessary, providing a faster experience. Finally, because our system is built on top of LLM capabilities, which are far from perfect, they could return sometimes buggy or broken code. Users had to either regenerate or manually debug to handle these failures. We expect these issues to improve as LLMs themselves do. As new foundational models are released, they should continually be inserted and experimented with within the system to measure effects on performance.

While traversing through the Design Matrix while using DynEx, users expressed a desire for simpler and more readable language. Future work can integrate LLM-summarization capabilities or other NLP techniques to improve clarity, succinctness, and cognitive load. In addition, DynEx currently does not allow users to continuously brainstorm additional features during implementation. It would be beneficial to help users identify what other features to explore once the initial prototype is complete, instead of the user being left to their own devices in improving the initial prototype.

%Blending design and implementation goes further than preventing users from having backtrack and making drastic changes. Rather, it should facilitate the creation of thoughtful and intuitive applications, and future work should continue to explore how LLMs can best support this process as their capabilities change.

%Although LLM's ability to generate code will undoubtedly improve, it is unclear how changes in new releases of foundational models will affect the bridge between design and implementation. Advances in LLMs may further enable design and implementation to be done concurrently

Code-generation abilities of LLMs will undoubtedly improve and accelerate exploratory programming; blending design exploration and implementation must also be aligned with these developments. Our work explores the role of the Design Matrix in understanding the problem space for exploratory programming. Future work should continue investigating other approaches to how LLMs can effectively bridge the gap between design and implementation and enhance the exploratory programming process.

\section{Conclusion}
The advancement of LLMs poses a unique opportunity for exploratory programming. LLMs can generate code rapidly and also help users enhance and refine their problem space. In this paper, we presented DynEx, a system to accelerate exploratory programming through a structured, LLM-guided Design Matrix and \revision{code-generation through Modular Stepwise Implementation.}
%We also introduce self-invoking multi-modal LLMs as a technique for LLM-based exploratory programming, allowing for faster creation of diverse and functional prototypes. 
Through our user evaluation with 10 technical experts, we found that DynEx's Design Matrix allowed to users to explore, refine, and realize feature-rich applications suitable for prototyping. We conclude by discussing how systems like DynEx can help bridge the gap between design and implementation, empowering a broad range of individuals to bring their ideas to life. 

%Ideas can be translated to reality. 

%Furthermore, LLMs can also be used to accelerate brainstorming processes themselves, bridging the gap between design and implementation.

\bibliographystyle{ACM-Reference-Format}
\bibliography{sample-base}

\newpage
\appendix
\section{Few-shot Examples for the Design Matrix}
\label{app:example_matrix}

\textbf{OKCupid}\\
\textit{PersonXIdea} \\ Single person \\
\textit{PersonXGrounding} \\ - Difficulty in finding potential partners who meet specific criteria like race, religion, age, or occupation.\\
- Challenges in efficiently filtering through dating apps to locate compatible matches.\\
- Need for an application that streamlines the process by allowing users to set precise criteria and receive curated suggestions.\\
\textit{ApproachXIdea} \\ Searchable Database to allow people to search for people based on specified criteria \\
\textit{ApproachXGrounding} \\ - Ensure the database includes comprehensive filters such as age, gender, sex, religion, and occupation to meet users' specific search criteria. \\ - Develop a robust and scalable search algorithm that efficiently handles large datasets and returns accurate results based on the selected filters. \\ - Implement user-friendly search and filtering interfaces that make it easy for users to apply multiple criteria and refine their search results. \\
\textit{InteractionXIdea}\\
Faceted Browsing \\
\textit{InteractionXGrounding} \\ - Provide users with multiple facet filters, including age, gender, religion, occupation, and location, to refine their search effectively. \\ - Ensure the UI dynamically updates search results in real-time as users adjust their facet filters, offering immediate feedback. \\ - Design intuitive navigation with clear options to reset filters, switch criteria, and save searches, using checkboxes, sliders, and dropdowns for ease of use.\\\\
\textbf{Tinder} \\ 
\textit{PersonXIdea}\\ Single person
\textit{PersonXGrounding}\\ - Users often struggle to find matches that meet their physical preferences quickly and efficiently. \\ - The abundance of profiles makes it challenging to identify those that align with specific looks or appearances. \\ - A streamlined approach to swiping and filtering by appearance would help users connect with potential matches faster, focusing on visual attraction.
\textit{ApproachXIdea} \\ Lower the cognitive load by providing less information, making it easier to judge potential matches quickly
\textit{ApproachXGrounding} \\ - Limit the displayed information to essential details, such as a single profile photo and a brief tagline, to encourage snap judgments. \\ - Focus on visual appeal as the primary matching criterion, reducing the need for users to sift through extensive profiles. \\ - Use an algorithm to prioritize matches based on visual preferences and minimal data inputs, streamlining the matching process.
\textit{InteractionXIdea} \\ Card Swipe
\textit{InteractionXGrounding} \\- Each card should prominently feature a large profile photo, as visual appeal is the primary factor in this interaction. \\ - Include minimal text, such as the person’s name, age, and a short tagline or fun fact, to provide just enough context without overwhelming the user. \\ - Add simple icons or buttons for actions like "Like" or "Pass," ensuring that users can quickly swipe or tap to make their choice. \\~\\
\textbf{Coffee Meets Bagel} \\ 
\textit{PersonXIdea} \\ Single person \\ 
\textit{PersonXGrounding} \\ - Users who are looking for serious relationships prefer fewer, high-quality matches over endless swiping. \\ - The overwhelming number of potential matches on other apps can make it difficult to focus on finding a meaningful connection.\\ - An app designed for serious dating should streamline the process by offering a curated selection of potential partners, reducing time spent on the app.
ApproachXIdea: Lower the cognitive load by having less matches to make more intentional judgements \\
\textit{ApproachXGrounding} \\- Present a select number of potential matches each day to prevent decision fatigue and promote thoughtful consideration.\\ - Display key information like shared interests, compatibility indicators, and mutual friends to aid decision-making without overwhelming the user. \\ - Prioritize quality over quantity, ensuring that each match is relevant to the user’s preferences and relationship goals. \\
\textit{InteractionXIdea} \\ Feed with 5 options to date
\textit{InteractionXGrounding} \\ - The daily message should include a concise profile summary for each of the five matches, highlighting essential details such as name, age, occupation, and a short personal note or shared interest. \\ - Include compatibility scores or commonalities (e.g., mutual friends, hobbies) to help users quickly assess each match’s potential. \\ - Provide clear action buttons within the message to either like, pass, or start a conversation, making it easy for users to engage with their daily options.

\section{Few-shot Examples for the Project Specification}
\label{app:example_spec}
\textit{Music Recommendation App} \\
Application Layout: \\
- Create a clean, simple interface divided into two main sections: "Discover" and "Favorites." \\
- The "Discover" section would have a large area that changes dynamically to display song details with each swipe (song name, artist, album, genre, and description). \\
- It would also have 'like', 'dislike' and 'skip' buttons, which will work with simple clicks. \\
- The 'Favorites' section would be a list of all liked songs. \\
User Interactions: \\
- The user can click to swipe a song left (dislike), right (like), or down (skip). \\
- The user can click on a song to save to favorites once liked.
- The user can navigate between "Discover" and "Favorites" sections via top navigation tabs. \\
Inputs and Logic: \\
- The app will use the user's interactions (likes, dislikes, skips) as inputs to an ML model (implemented using GPT) to evolve its music recommendations in real-time. \\
- On initial use, ask the user for favorite genres, artists, or songs to kickstart the ML algorithm. \\
- The user's interaction with each song (whether they like, dislike, or skip it) will further tailor the recommendations. \\
- The saved favorite songs will be stored \\
- There is no need to create placeholder data for music, as GPT will return the music recommendations. \\~\\
\textit{Outfit Generator App} \\
Application Layout: \\
- Create a clean, minimalist interface with a prominent central area for displaying outfit recommendations. \\
- Divide the interface into three main sections: "Outfit Recommendations," "Wardrobe," and "Saved Outfits." \\
- The "Outfit Recommendations" section should display swipeable cards with visual representations of the recommended outfits, along with relevant tags (season, occasion, style). \\
- The "Wardrobe" section should allow users to input their clothing items, categorized by type (tops, bottoms, dresses, etc.). \\
- The "Saved Outfits" section should display a grid of liked outfits for future reference. \\
User Interactions: \\
- Users can swipe left by clicking no, or right by clicking yes on the outfit recommendation cards to dislike or like the outfit, respectively. \\
- Users can click on individual clothing items in the "Wardrobe" section to add or remove them from their virtual wardrobe. \\
- Users can click on a liked outfit in the "Saved Outfits" section to view its details or remove it from the saved list. \\
Inputs and Logic: \\
- The app will use the user's initial wardrobe inputs and style preferences (gathered through a brief questionnaire) to kickstart the GPT-powered outfit recommendation algorithm. \\
- The algorithm will consider factors like season, occasion, and the user's wardrobe items to generate outfit recommendations. \\
- The user's interactions (likes, dislikes) with the recommended outfits will be used as feedback to refine and personalize the algorithm's recommendations over time. \\
- The liked outfits will be saved in the "Saved Outfits" section for future reference. \\
- Create placeholder data for the initial wardrobe. \\~\\
\textit{Plant Watering Tracker} \\
Application Layout: \\
- Create a clean and intuitive interface with a prominent section for the "Watering Calendar." \\
- Divide the interface into three main sections: "Watering Calendar," "Plant List," and "Watering Reminders." \\
- The "Watering Calendar" section should display a monthly calendar view with visual indicators for scheduled watering days. \\
- The "Plant List" section should allow users to add and manage their plant collection, including details like species, pot size, and watering requirements. \\
- The "Watering Reminders" section should display upcoming watering tasks and allow users to set notification preferences. \\
User Interactions: \\
- Users can click on specific dates in the "Watering Calendar" to schedule or modify watering tasks for individual plants or groups. \\
- Users can click on plants in the "Plant List" to view or edit their details, including watering schedules. \\
- Users can set notification preferences (email, push notifications, etc.) for upcoming watering tasks in the "Watering Reminders" section. \\
Inputs and Logic: \\
- The app will use the user's initial plant inputs (species, pot size, etc.) to determine baseline watering requirements for each plant. \\
- An algorithm (implemented using GPT) will analyze factors like plant type, pot size, and environmental conditions (temperature, humidity, etc.) to generate adaptive watering schedules. \\
- The algorithm will learn from the user's interactions (manually adjusting watering schedules, plant health feedback) to refine its recommendations over time. \\
- The app will send reminders based on the user's scheduled watering tasks and notification preferences. \\
- Create placeholder data for the user's current plants. \\
\section{Prompt and Few-shot Examples for Placeholder Data}
\label{app:example_data}
You are generating fake JSON data for a UI that a user wants to create. The spec should give instructions as to what data needs to be generated.

        For example, for this spec:\\
        "Application Layout:\\
        - Create a clean, minimalist interface with a prominent central area for displaying outfit recommendations.\\
        - Divide the interface into three main sections: "Outfit Recommendations," "Wardrobe," and "Saved Outfits."\\
        - The "Outfit Recommendations" section should display swipeable cards with visual representations of the recommended outfits, along with relevant tags (season, occasion, style).\\
        - The "Wardrobe" section should allow users to input their clothing items, categorized by type (tops, bottoms, dresses, etc.).\\
        - The "Saved Outfits" section should display a grid of liked outfits for future reference.\\
        User Interactions:\\
        - Users can swipe left by clicking no, or right by clicking yes on the outfit recommendation cards to dislike or like the outfit, respectively.\\
        - Users can click on individual clothing items in the "Wardrobe" section to add or remove them from their virtual wardrobe.\\
        - Users can click on a liked outfit in the "Saved Outfits" section to view its details or remove it from the saved list.\\
        Inputs and Logic:\\
        - The app will use the user's initial wardrobe inputs and style preferences (gathered through a brief questionnaire) to kickstart the GPT-powered outfit recommendation algorithm.\\
        - The algorithm will consider factors like season, occasion, and the user's wardrobe items to generate outfit recommendations.\\
        - The user's interactions (likes, dislikes) with the recommended outfits will be used as feedback to refine and personalize the algorithm's recommendations over time.\\
        - The liked outfits will be saved in the "Saved Outfits" section for future reference.\\
        - Create placeholder data for initial wardrobe."
  we only want to generate fake data for the INITIAL wardrobe, not the outfit recommendations.

  Also, consider the user in this situation, and generate data tailored to the user if necessary. The application is for {person\_idea}, with these details: {person\_grounding}

	Please generate a JSON array of fake data with appropriate fields. Here is an example:

        Input: I want to create a UI that visualizes a beauty store's inventory... It should have the fields `title`, `description`, `price`, `discountPercentage`, `rating`, `stock`, `brand`, `category`

        System result:
        \begin{verbatim}
        [
            {{
                "id": 11,
                "title": "perfume Oil",
                "description": "Mega Discount",
                "price": 13,
                "discountPercentage": 8.4,
                "rating": 4.26,
                "stock": 65,
                "brand": "Impression of Acqua Di Gio",
                "category": "fragrances",
            }},
            {{
                "id": 12,
                "title": "perfume Oil",
                "description": "Half Off",
                "price": 15,
                "discountPercentage": 12.3,
                "rating": 3.46,
                "stock": 2343,
                "brand": "Victoria Secret",
                "category": "fragrances",
            }},
        ]
        \end{verbatim}
        Please follow these rules while creating the JSON array \\
        1. Please only return the JSON array and nothing else. \\
        2. Array length should be length 10-20. \\
        3. Please ensure that the generated data makes sense. \\

\section{Few-shot Examples for a Stepwise Plan Given Spec}
\label{app:example_plan}
\textbf{Spec} \\
Application Layout\\
- Create a clean, minimalist interface with a prominent central area for displaying outfit recommendations.\\
- Divide the interface into three main sections: "Outfit Recommendations," "Wardrobe," and "Saved Outfits."\\
- The "Outfit Recommendations" section should display swipeable cards with visual representations of the recommended outfits, along with relevant tags (season, occasion, style).\\
- The "Wardrobe" section should allow users to input their clothing items, categorized by type (tops, bottoms, dresses, etc.).\\
- The "Saved Outfits" section should display a grid of liked outfits for future reference.\\
User Interactions:\\
- Users can swipe left by clicking no, or right by clicking yes on the outfit recommendation cards to dislike or like the outfit, respectively.\\
- Users can click on individual clothing items in the "Wardrobe" section to add or remove them from their virtual wardrobe.\\
- Users can click on a liked outfit in the "Saved Outfits" section to view its details or remove it from the saved list.\\
Inputs and Logic:\\
- The app will use the user's initial wardrobe inputs and style preferences (gathered through a brief questionnaire) to kickstart the GPT-powered outfit recommendation algorithm.\\
- The algorithm will consider factors like season, occasion, and the user's wardrobe items to generate outfit recommendations.\\
- The user's interactions (likes, dislikes) with the recommended outfits will be used as feedback to refine and personalize the algorithm's recommendations over time.\\
- The liked outfits will be saved in the "Saved Outfits" section for future reference.\\
- Create placeholder data for the initial wardrobe.\\

\textbf{Plan}\\
1. Set up the React application and create the main layout with the three sections: 'Outfit Recommendations', 'Wardrobe', and 'Saved Outfits'. Read in the placeholder data from the endpoint.\\
2. Implement the 'Outfit Recommendations' section with swipeable cards using MUI components. Create placeholder data for initial outfit recommendations.\\
3. Implement the 'Wardrobe' section with a list of clothing items categorized by type (tops, bottoms, dresses, etc.). Allow users to add or remove items from their virtual wardrobe.\\
4. Implement the 'Saved Outfits' section with a grid layout to display liked outfits. Allow users to view outfit details or remove outfits from the saved list.\\
5. Integrate GPT to generate outfit recommendations based on the user's wardrobe and style preferences. Implement the logic to handle user interactions (likes, dislikes) and refine the recommendations accordingly.\\

\section{Prompt for a Basic Code Generation}
\label{app:example_code_basic}
The entire app will be in one index.html file. It will be written entirely in HTML, Javascript, and CSS. The design should not incorporate routes. Everything should exist within one page. No need for design mockups, wireframes, or external dependencies.\\
The entire app will be written using React and MUI. Load MUI from the CDN. Here is an example: \\
\begin{verbatim}
    <!DOCTYPE html>
<html lang="en">
<head>
  <meta charset="UTF-8">
  <meta name="viewport" content="width=device-width,
  initial-scale=1.0">
  <title>React App with MUI and Hooks</title>
  <!-- Load React and ReactDOM from CDN -->
  <script src="https://unpkg.com/react@18/umd/
  react.development.js" crossorigin></script>
  <script src="https://unpkg.com/react-dom@18/
  umd/react-dom.development.js" crossorigin></script>
  <!-- Babel for JSX transformation -->
  <script src="https://unpkg.com/@babel/standalone/
  babel.min.js"></script>
  <!-- Load MUI from CDN -->
  <link rel="stylesheet" href="https://fonts.googleapis
  .com/css?family=Roboto:300,400,500,700&display=swap" />
  <script src="https://unpkg.com/@mui/material@
  5.0.0-rc.1/umd/material-ui.development.js" crossorigin>
  </script>
</head>
<body>
  <div id="root"></div>
  <script type="text/babel">
    const {
      Button,
      Container,
      Typography,
      TextField,
    } = MaterialUI;

    const { useState, useEffect } = React;

    function App() {
      const [count, setCount] = useState(0);
      const [name, setName] = useState('');

      useEffect(() => {
        document.title = \`Count: \${count}\`;
      }, [count]);

      return (
        <Container>
          <Typography variant="h2" component="h1" 
          gutterBottom>
            Hello, React with Material-UI and Hooks!
          </Typography>
          <Typography variant="h5">
            Count: {count}
          </Typography>
          <Button variant="contained" color="primary" 
          onClick={() => setCount(count + 1)}>
            Increment
          </Button>
          <TextField
            label="Name"
            value={name}
            onChange={(e) => setName(e.target.value)}
            variant="outlined"
            margin="normal"
            fullWidth
          />
          <Typography variant="h6">
            Name: {name}
          </Typography>
        </Container>
      );
    }

    const rootElement = document.getElementById('root');
    const root = ReactDOM.createRoot(rootElement);
    root.render(<App />);
  </script>
</body>
</html>
\end{verbatim}
- DO NOT DELETE PREVIOUS CODE. DO NOT RETURN A CODE SNIPPET. RETURN THE ENTIRE CODE. Only ADD to existing code to implement the task properly. DO NOT COMMENT PARTS OF THE CODE OUT AND WRITE /*...rest of the code */ or something similar. DO NOT COMMENT ANY PARTS OF THE CODE OUT. DO NOT COMMENT ANY PARTS OF THE CODE OUT FROM PREVIOUS CODES.\\
- DO NOT COMMENT {{/* Other sections remain the same */}}. ALL THE CODE MUST EXIST. ALL YOU ARE DOING IS ADDING FUNCTIONALITY. ADDING FUNCTIONALITY - DO NOT REMOVE ANY PREVIOUS CODE.\\
- The entire file should be less than 420 lines of code. MUST BE LESS THAN 420 LINES OF CODE.\\
- DO NOT LOAD ANYTHING ELSE IN THE CDN. Specifically, DO NOT USE: MaterialUI Icon, Material UI Lab.\\
- Do not return separate code files. All the components should be in one code file and returned.\\
- Do not type import statements. Assume that MUI and react are already imported libraries, so to use the components simply do so like this: const \{{Button, Container, Typography, TextField \}} = MaterialUI; or const \{{ useState, useEffect \}} = React;\\
- DO NOT USE THESE MUI COMPONENTS: Calendar, DatePicker, TimePicker, SwipeableViews, SwipeableViewsVirtualizer, Fade, MopbileStepper as they do not exist.\\

\section{Few-shot Example for Self-Invoking GPT}
\label{app:example_gpt}
\begin{verbatim}
try {{
 const response = await fetch('https://api.openai.com/v1
 /chat/completions', {{
   method: 'POST',
   headers: {{
     'Content-Type': 'application/json',
     'Authorization': `Bearer {openai_api_key}`
   }},
   body: JSON.stringify({{
     model: 'gpt-4',
     messages: [
       {{
         role: 'system',
         content: 'You are a helpful assistant providing 
         clothing recommendations based on user preferences.'
       }},
       {{
         role: 'user',
         content: `
         Based on the following preferences, provide a list 
         of recommended clothing items in the following JSON 
         format:
         [
           {{
             "id": 6,
             "itemName": "Striped T-Shirt",
             "description": "A classic striped t-shirt made 
             of 100% cotton.",
             "imageUrl": "https://example.com/striped-
             tshirt.jpg",
             "size": "M",
             "brand": "H&M",
             "price": 19.99
           }},
           ...
         ]

         Preferences: XYX

         Ensure the JSON is valid and adheres strictly to 
         this format. Do not type any additional text, 
         only provide the JSON.`
       }}
     ]
   }})
 }});


 const result = await response.json();
 const parsedData = JSON.parse(result.choices[0].message.
 content);
 setRecommendations(parsedData);
}} catch (err) {{
 setError(err.message);
}} finally {{
 setLoading(false);
}}
\end{verbatim}
\section{Few-shot Example for Self-Invoking OpenAI for Images}
\label{app:example_images}
\begin{verbatim}
    try:
   response = await fetch('https://api.openai.com/v1/chat
   /completions', {{
       'method': 'POST',
       'headers': {{
           'Content-Type': 'application/json',
           'Authorization': 'Bearer {openai_api_key}'
       }},
       'body': json.dumps({{
           'model': 'gpt-4',
           'messages': [
               {{
                   'role': 'system',
                   'content': 'You are a helpful assistant 
                   providing grocery recommendations based on 
                   dietary preferences and restrictions.'
               }},
               {{
                   'role': 'user',
                   'content': \"\"\"
                   Based on the following dietary preferences
                   and restrictions, provide a list of 
                   recommended grocery items in the following 
                   JSON format:
                   [
                       {{
                           "name": "Organic Quinoa",
                           "description": "A gluten-free, 
                           high-protein grain.",
                           "category": "Grains",
                           "nutritionalInfo": {{
                               "calories": 120,
                               "fat": 2.1,
                               "protein": 4.4,
                               "carbs": 21.3
                           }},
                           "compatibility": "Vegan, Gluten-Free, 
                           Vegetarian"
                       }},
                       ...
                   ]


                   Preferences and Restrictions: {{preferences}}


                   Ensure the JSON is valid and adheres strictly
                   to this format. Do not type any additional
                   text, only provide the JSON.
                   \"\"\"
               }}
           ]
       }})
   }})


   result = await response.json()
   parsed_data = json.loads(result['choices'][0]['message']
   ['content'])


   # Fetch images for each recommendation
   for item in parsed_data:
       image_response = await fetch('https://api.openai.com/
       v1/images/generations', {{
           'method': 'POST',
           'headers': {{
               'Content-Type': 'application/json',
               'Authorization': 'Bearer {openai_api_key}'
           }},
           'body': json.dumps({{
               'prompt': f"An image of {{item['name']}}, 
               a {{item['category']}} item.",
               'n': 1,
               'size': "256x256"
           }})
       }})


       image_result = await image_response.json()
       item['imageUrl'] = image_result['data'][0]['url']


   set_recommendations(parsed_data)


except Exception as err:
   print(err)
   set_error(str(err))


finally:
   set_loading(False)

\end{verbatim}

\section{Prompt for Iteration Within Steps to Debug}
\label{app:example_debug}
\textit{User Message}\\
Please fix the problem that the user describes: \color{blue}\{problem\} \color{black}\\
\textit{System Message}\\~\\
A coding task has been implemented for a project we are working on. For context, this is the project description: \color{blue}\{spec\}\color{black}. The task was this: \color{blue}\{task
\}\color{black}. This is the faked\_data: \color{blue}\{faked\_data\}\color{black}. However, the task was not implemented fully correctly. The user explains what is wrong in the problem \color{blue}\{problem\}\color{black}. There is already existing code in the index.html file. Using the existing code \color{blue}\{task\_code\}\color{black}. Please fix the problem.
PLEASE DO NOT DELETE EXISTING CODE. ONLY FIX THE BUG.
Return the FULL CODE NEEDED TO HAVE THE APP WORK, INSIDE THE INDEX.HTML file.

\section{Design Matrix for Usage Scenario}
\label{app:usage_matrix}
\textbf{Person:Idea} \\
\begin{tabular}{c c c} 
    \fbox{
        \parbox[c][2cm][c]{0.12\textwidth}{
            \centering 1. Non-native speakers interested in Chinese culture
        }
    } & 
    \setlength{\fboxrule}{2pt} % Set the border thickness for the middle box
    \fbox{
        \parbox[c][2cm][c]{0.12\textwidth}{
            \centering 2. Visual learners struggling with language memorization
        }
    } 
    \setlength{\fboxrule}{0.5pt} % Reset the border thickness for subsequent boxes
    & 
    \fbox{
        \parbox[c][2cm][c]{0.12\textwidth}{
            \centering 3. Travel enthusiasts planning a trip to China
        }
    } \\
    \fbox{
        \parbox[c][2cm][c]{0.12\textwidth}{
            \centering 4. Retired adult wanting to expand linguistic skills
        }
    } & 
    \fbox{
        \parbox[c][2cm][c]{0.12\textwidth}{
            \centering 5. Busy university student wanting to study Chinese in his free time
        }
    } & 
    \fbox{
        \parbox[c][2cm][c]{0.12\textwidth}{
            \centering 6. Travel enthusiasts planning a trip to China
        }
    }
\end{tabular}\\~\\~\\
\textbf{Person:Grounding} \\
\begin{tabular}{c c c} 
    \fbox{
        \parbox[c][2.5cm][c]{0.45\textwidth}{
            \centering  - Confusion arises due to unfamiliarity with Chinese characters and their complex structure, slowing the learning process. \\
            - Difficulty in linking characters to their corresponding meaning or pronunciation, hindering vocabulary acquisition. \\
            - Traditional memorization methods offer little aid to visual learners who could better recall information through imagery.
        }
    }
\end{tabular}\\~\\~\\
\textbf{Approach:Idea (1)} \\
\begin{tabular}{c c c} 
    \setlength{\fboxrule}{2pt} % Set the border thickness for the middle box
    \fbox{
        \parbox[c][2cm][c]{0.12\textwidth}{
            \centering 1. Pictorial spaced repetition learning
        }
    }     \setlength{\fboxrule}{0.5pt} % Reset the border thickness for subsequent boxes
& 
    \fbox{
        \parbox[c][2cm][c]{0.12\textwidth}{
            \centering 2. Visual storytelling for language acquisition
        }
    } 
    & 
    \fbox{
        \parbox[c][2cm][c]{0.12\textwidth}{
            \centering 3. Cognitive load theory for efficient memorization
        }
    } 
\end{tabular}\\~\\~\\
\textbf{Approach:Grounding (1)} \\
\begin{tabular}{c c c} 
    \fbox{
        \parbox[c][5cm][c]{0.45\textwidth}{
            \centering  - Implement the spaced repetition system (SRS) algorithm to schedule review times according to each user’s progress, allowing items to reappear before they’re likely to be forgotten. \\
            - Incorporate visual  images representing Chinese characters or words, enhancing the memory association between visual cues and corresponding meanings. \\
            - Integrate regular reviews into the learning process, which are fine-tuned according to the user’s performance, to further solidify memory and retention. \\
            - Leverage cognitive science principles such as interleaving (mixing similar tasks) and retrieval practice (recalling an item from memory), to increase learning effectiveness. 
        }
    }
\end{tabular}\\~\\~\\
\textbf{Approach:Idea (2)} \\
\begin{tabular}{c c c} 
    \fbox{
        \parbox[c][2cm][c]{0.12\textwidth}{
            \centering 1. Pictorial spaced repetition learning
        }
    } & 
    \setlength{\fboxrule}{2pt} % Set the border thickness for the middle box
    \fbox{
        \parbox[c][2cm][c]{0.12\textwidth}{
            \centering 2. Visual storytelling for language acquisition
        }
    } 
    \setlength{\fboxrule}{0.5pt} % Reset the border thickness for subsequent boxes
    & 
    \fbox{
        \parbox[c][2cm][c]{0.12\textwidth}{
            \centering 3. Cognitive load theory for efficient memorization
        }
    } 
\end{tabular}\\~\\~\\
\textbf{Approach:Grounding (2)} \\
\begin{tabular}{c c c} 
    \fbox{
        \parbox[c][4.5cm][c]{0.45\textwidth}{
            \centering  - Integrate visual aids, such as illustrations or animations, that correlate with each word’s meaning to enhance understanding and recall. \\
            -   Develop an algorithm that links related words and images together in a meaningful story, promoting stronger memory associations.\\
            - Make use of GPT to generate context-rich sentences or mini-stories, helping to create a narrative around each word or character. \\ 
            - Ensure design of the learning material caters to visual learners with a focus on vibrant, engaging, and contextually relevant graphical representations. 
        }
    }
\end{tabular}\\~\\~\\
\textbf{Interaction:Idea} \\
\begin{tabular}{c c c} 
    \setlength{\fboxrule}{2pt} % Set the border thickness for the middle box
    \fbox{
        \parbox[c][2cm][c]{0.12\textwidth}{
            \centering 1. Simple guess-and-review quiz interface 
        }
    } \setlength{\fboxrule}{0.5pt} % Reset the border thickness for subsequent boxes
& 
    \fbox{
        \parbox[c][2cm][c]{0.12\textwidth}{
            \centering 2. Visual dictionary flashcard interface
        }
    } 
    & 
    \fbox{
        \parbox[c][2cm][c]{0.12\textwidth}{
            \centering 3. Image-based language learning game interface
        }
    } 
\end{tabular}\\~\\~\\
\textbf{Interaction:Grounding} \\
\begin{tabular}{c c c} 
    \fbox{
        \parbox[c][3.5cm][c]{0.45\textwidth}{
            \centering  - The quiz interface should present the Chinese character or word along with its corresponding image. \\
            - The user then attempts to guess its meaning. If they respond accurately, the item is pushed back into the review cycle based on the SRS algorithm. \\
            - If the guess is incorrect, the correct meaning is displayed, and the item is scheduled for another review sooner. \\
            - Users should have a clear view of their progress and a way to navigate to previously learned words for self-study. 
        }
    }
\end{tabular}\\~\\

\section{Spec for Usage Scenario}
\label{app:usage_spec}
Application Layout:\\
- Divide the interface into three main sections: "Flashcards," "Quiz," and "Progress."\\
- The "Flashcards" section displays the Chinese character/word along with its corresponding image for visual association.\\
- The "Quiz" section presents the character/word and image, prompting the user to input the meaning.\\
- The "Progress" section shows the user's performance metrics, such as accuracy rate and words mastered.\\
\\~\\
User Interactions:\\
- In the "Flashcards" section, users can click through the flashcards to study the character/word and its associated image.\\
- In the "Quiz" section, users can type in their answer for the displayed character/word and image.\\
- Users can click a "Show Answer" button to reveal the correct meaning if they are unsure.\\
- Users can navigate between the "Flashcards," "Quiz," and "Progress" sections using tabs or buttons.\\
\\~\\
Inputs and Logic:\\
- The app will use a pre-generated dataset of Chinese characters/words, their meanings, and associated images.\\
- The spaced repetition system (SRS) algorithm will determine the optimal interval for reviewing each character/word based on the user's performance.\\
- Correctly answered items in the "Quiz" section will have their review interval increased (pushed further back in the review cycle).\\
- Incorrectly answered items will have their review interval decreased (scheduled for more frequent review).\\
- The app will track the user's progress, such as the number of words mastered and accuracy rates, and display this information in the "Progress" section.\\
- Incorporate cognitive science principles like interleaving (mixing different types of content) and retrieval practice (actively recalling information) into the "Quiz" section to enhance learning effectiveness.\\
\\
By combining visual associations with images, spaced repetition, and interactive quizzes, this UI aims to provide an engaging and effective way for non-Chinese speaking students to learn and retain Chinese characters and vocabulary.
\section{Placeholder Data for Usage Scenario}
\label{app:usage_data}
\begin{CJK}{UTF8}{gbsn}  % Use the CJK environment with the correct font
\begin{verbatim}
    [
  {
    "id": 1,
    "chinese": "我学习中文",
    "meaning": "I study Chinese",
    "imagePath": "images/studying.jpg"
  },
  {
    "id": 2,
    "chinese": "你好吗?",
    "meaning": "How are you?",
    "imagePath": "images/greeting.jpg"
  },
  {
    "id": 3,
    "chinese": "这是一本书",
    "meaning": "This is a book",
    "imagePath": "images/book.jpg"
  },
  {
    "id": 4,
    "chinese": "我喜欢吃苹果",
    "meaning": "I like to eat apples",
    "imagePath": "images/apple.jpg"
  },
  {
    "id": 5,
    "chinese": "今天天气很好",
    "meaning": "The weather is nice today",
    "imagePath": "images/sunny.jpg"
  },
  {
    "id": 6,
    "chinese": "我们去公园吧",
    "meaning": "Let's go to the park",
    "imagePath": "images/park.jpg"
  },
  {
    "id": 7,
    "chinese": "这个房间很大",
    "meaning": "This room is big",
    "imagePath": "images/room.jpg"
  },
  {
    "id": 8,
    "chinese": "我喜欢听音乐",
    "meaning": "I like to listen to music",
    "imagePath": "images/music.jpg"
  },
  {
    "id": 9,
    "chinese": "我们去看电影吧",
    "meaning": "Let's go watch a movie",
    "imagePath": "images/movie.jpg"
  },
  {
    "id": 10,
    "chinese": "我想买一件新衣服",
    "meaning": "I want to buy a new piece of clothing",
    "imagePath": "images/shopping.jpg"
  },
  {
    "id": 11,
    "chinese": "我们去吃晚饭吧",
    "meaning": "Let's go eat dinner",
    "imagePath": "images/dinner.jpg"
  },
  {
    "id": 12,
    "chinese": "这个城市很繁华",
    "meaning": "This city is bustling",
    "imagePath": "images/city.jpg"
  },
  {
    "id": 13,
    "chinese": "我喜欢旅游",
    "meaning": "I like to travel",
    "imagePath": "images/travel.jpg"
  },
  {
    "id": 14,
    "chinese": "我们去运动吧",
    "meaning": "Let's go exercise",
    "imagePath": "images/exercise.jpg"
  },
  {
    "id": 15,
    "chinese": "这个项目很有趣",
    "meaning": "This project is interesting",
    "imagePath": "images/project.jpg"
  }
]
\end{verbatim}
\end{CJK}
\section{Plan for Usage Scenario}
\begin{enumerate}
\item{Set up the React application and create the main layout with the three sections: 'Flashcards', 'Quiz', and 'Progress'. Read in the pre-generated dataset of Chinese characters/words, their meanings, and associated images from the data endpoint.}
\item{Implement the 'Flashcards' section with a display for the Chinese character/word and its corresponding image. Allow users to click through the flashcards.}
\item{Implement the 'Quiz' section with a display for the Chinese character/word and its associated image. Allow users to input their answer and reveal the correct meaning. Integrate the spaced repetition system (SRS) algorithm to determine the optimal review interval for each character/word based on the user's performance.}
\item{Implement the 'Progress' section to display the user's performance metrics, such as accuracy rate and words mastered. Incorporate cognitive science principles like interleaving and retrieval practice into the 'Quiz' section.}
\item{Call GPT to generate images for the Chinese characters/words and associate them with the corresponding entries in the dataset. Display these images in the 'Flashcards' and 'Quiz' sections.}
\end{enumerate}

\end{document}